\documentclass[aps,prb,reprint,superscriptaddress]{revtex4-2}
\usepackage[colorlinks,linkcolor={blue},citecolor={blue},urlcolor={blue}]{hyperref} 
\usepackage{graphicx}
\usepackage{subfigure}
\usepackage{amsmath}
\usepackage{braket}
\usepackage{soul} 
\usepackage[T1]{fontenc}
\usepackage[utf8]{inputenc}
\usepackage{float}
\usepackage{marvosym} 
\usepackage{wasysym} 
\usepackage{bm}
\usepackage{amssymb} 
\usepackage[normalem]{ulem}
\usepackage{esint} 

\makeatletter
\def\maketitle{
\@author@finish
\title@column\titleblock@produce
\suppressfloats[t]}
\makeatother
\newcommand{\bk}{\boldsymbol{k}}
\newcommand{\bs}{\boldsymbol}
\newcommand{\Cal}{\mathcal}

\newcommand{\bA}{\boldsymbol{A}}

\begin{document}



\title{Fractional quantum anomalous Hall effect in a singular flat band}


\author{Wenqi Yang}
\thanks{Wenqi Yang and Dawei Zhai contributed equally to this work.}
\affiliation{New Cornerstone Science Laboratory, Department of Physics, The University of Hong Kong, Hong Kong, China}
\author{Dawei Zhai}
\thanks{Wenqi Yang and Dawei Zhai contributed equally to this work.}
\affiliation{New Cornerstone Science Laboratory, Department of Physics, The University of Hong Kong, Hong Kong, China}
\author{Tixuan Tan}
\affiliation{Department of Physics, Stanford University, Stanford, CA 94305, USA}
\author{Feng-Ren Fan}
\affiliation{New Cornerstone Science Laboratory, Department of Physics, The University of Hong Kong, Hong Kong, China}
\author{Zuzhang Lin}
\affiliation{New Cornerstone Science Laboratory, Department of Physics, The University of Hong Kong, Hong Kong, China}
\author{Wang Yao}
\email{wangyao@hku.hk}
\affiliation{New Cornerstone Science Laboratory, Department of Physics, The University of Hong Kong, Hong Kong, China}



\begin{abstract}

In the search of fractional quantum anomalous Hall (FQAH) effect, the conventional wisdom is to start from a flat Chern band isolated from the rest of the Hilbert space by band gaps, so that many-body interaction can be projected to a landscape that mimics a Landau level. Singular flat bands (SFB), which share protected touching points with other dispersive bands, represent another type of flat landscapes differing from Landau levels and Chern bands in topological and geometric properties. Here we report the finding of FQAH phases in a SFB, which emerges in the bipartite limit of the nearest-neighbor tight-binding model of twisted bilayer MoTe$_2$. 
At 1/3 and 2/3 filling of the SFB, FQAH effects are demonstrated using density matrix renormalisation group calculations with all bands, as well as exact diagonalization calculations with the two touching bands. Gapping the band touching can turn the SFB into a nearly flat Chern band, but counter-intuitively this suppresses the FQAH effect, as the gap opening introduces strong inhomogeneity to the quantum geometry. An optical scheme to realize such SFB for cold atoms is provided. Our findings uncover a new arena for the exploration of fractional quantum Hall physics beyond the Landau level and Chern insulator paradigms.

\end{abstract}

\pacs{}

\maketitle



Quantum anomalous Hall effects, the zero-magnetic-field analog of the quantum Hall effects, are widely explored in magnetic topological insulators~\cite{Haldane1988,Tokura2019,chang2023colloquium}. These insulators host Bloch bands of nonzero Chern numbers that resemble Landau levels in an applied magnetic field~\cite{TKNN1982}, leading to the observations of integer quantum anomalous Hall effects~\cite{chang2013experimental,YuanboScience2020,AndreaYoungScience2020,LiTingxinNature2021}. 
Besides topology, another important aspect of Landau levels is their flatness in dispersion and quantum geometry, as well as isolation by energy gaps. This underlies the emergence of fractional quantum Hall effects when many-body interaction projected to individual Landau levels dominates the electron correlation at fractional fillings~\cite{Halperin2020fractional}. 
The search for fractional quantum anomalous Hall (FQAH) effect has therefore naturally started from insulators having a flat Chern band well isolated from all other bands by energy gaps. There has been a remarkable surge of research to engineer such band Chern insulators in various lattice models~\cite{tang2011high, wu2012zoology,sun2011nearly, neupert2011fractional, sheng2011fractional, wang2011fractional, regnault2011fractional,wu2012zoology,xiao2011interface,venderbos2012fractional,wu2012zoology},
where FQAH phases (or fractional Chern insulators~\cite{regnault2011fractional}) are revealed under short-range interactions by exact diagonalization (ED) or density matrix renormalization group (DMRG) calculations. 
Most notably, experimental observations of FQAH effects are recently reported in the isolated Chern band of twisted rhombohedral bilayers of MoTe$_2$~\cite{FCIMoTe2Jiaqi2023,FCIMoTe2ShanJie2023,FCIMoTe2Park2023,FCIMoTe2PRX2023} and pentalayer graphene/hBN moir\'e~\cite{lu2024fractional}, which are significantly advancing the research field.

The emergence of flat topological band is subject to fundamental constraints. In 2D lattices of finite-ranged hopping, a Bloch band can not simultaneously possess exact flatness, nonzero Chern number, and isolation by energy gap~\cite{FlatChernTheorem}. In a variety of lattice models, such constraint underlies another intriguing context for the exploration of strong correlation and topological physics, i.e. \textit{singular flat band} (SFB), where exact band flatness is protected by point touching with dispersive band(s) ~\cite{BandTouchingBalents2008,SinguFlatClassificationPRB2019,SinguFlatAdvPhysX2021}. These lattices host two types of states localized by destructive hopping interference~\cite{BandTouchingBalents2008}: compact localized states of 0D character; and 1D states forming non-contractible loops encircling the torus geometry under periodic boundary condition. Contribution of the non-contractible loop states to the state counting at the flat-band energy dictates the presence of the point touching as immovable singularity of the band~\cite{BandTouchingBalents2008,SinguFlatClassificationPRB2019,SinguFlatAdvPhysX2021}.
In applied magnetic fields, a SFB can develop Landau levels at unexpected energies, and exhibit diverging orbital magnetic susceptibility~\cite{SinguFlatNature2020}.
Along with these exotic single-particle properties, the exact band flatness offers prospects for strong correlation physics. However, the band touching is often considered to be troublesome, rendering the projection onto the flat band problematic and band Chern number undefinable.

\begin{figure}[t]
	\centering
	\includegraphics[width=3.4in]{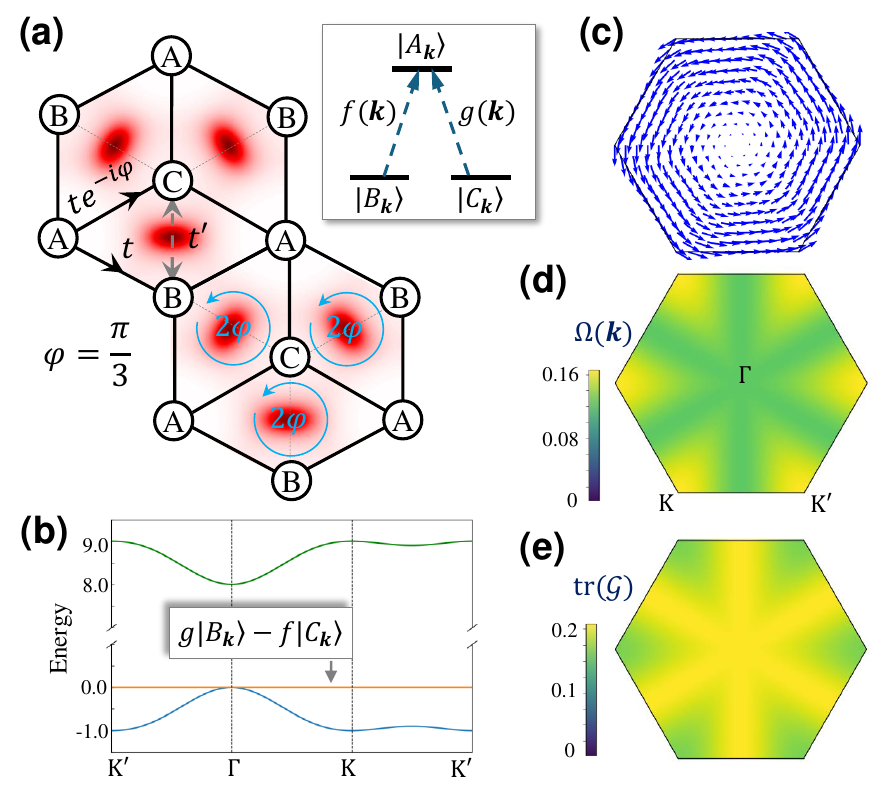}
	\caption{(a) Schematics of the dice lattice, with red color in the background denoting a pseudo-magnetic field (see Fig.~\ref{Fig:System}). Inset: at an arbitrary wave vector $\bs{k}$, coupling between the three orbitals realizes a $\Lambda$ scheme. (b) Band structure in the non-interacting limit. (c) Under a non-periodic atomic gauge (see text), Berry connection of the SFB is continuous at the touching point $\Gamma$. (d) Berry curvature $\Omega$ and (e) trace of quantum metric tensor $\Cal{G}$ of the SFB. Here $E_A=8$, and the lattice constant is set to 1. All energies here and hereafter are measured in units of the hopping amplitude $t$.  
}
	\label{Fig1}
\end{figure}

In this work we report the finding of FQAH states in a SFB, which emerges in the bipartite limit of the nearest-neighbor tight-binding model of twisted bilayer MoTe$_2$~\cite{HongyiNSR2020}.
At 1/3 and 2/3 fillings of the SFB, FQAH effects are demonstrated with real-space DMRG calculations taking into account all bands with nearest-neighbor repulsive interaction ($U_1$). 
ED calculations on the two touching bands corroborate the DMRG results and reveal that the FQAH states are predominantly weighted in the SFB. 
For FQAH phases at both fillings, lower bound in $U_1$ is not observed within our computation capacity, consistent with the exact band flatness. Analysis of the ground state wave functions, as well as Hartree-Fock bands at the closest integer fillings, suggests that interaction-driven gap opening of the band touching is the cause for the transition from FQAH to a topologically trivial phase with the increase of $U_1$, which occurs at very different $U_1$ magnitudes at $1/3$ and $2/3$ SFB fillings. Gapping the band touching turns the SFB into a nearly flat isolated Chern band, but counter-intuitively this destroys the FQAH state as the gap opening introduces strong inhomogeneity to the quantum geometry. We also present a scheme to realize such SFB in optical lattice.

{\bf SFB model and its quantum geometry} -- A minimal model for realizing SFBs with nearest-neighbor hopping is a three-orbital one with the Bloch Hamiltonian (see Supplemental Material~\cite{supp} Sec.~S1)
\begin{equation}
    \hat{H}_0(\bk) = E_A\,\hat{A}^{\dagger}_{\bk} \hat{A}_{\bk} + f(\bk) \hat{A}^{\dagger}_{\bk} \hat{B}_{\bk} + g(\bk) \hat{A}^{\dagger}_{\bk} \hat{C}_{\bk} +h.c..~\label{eq:H0_k_space}
\end{equation}
This generic model has a flat band at the zero energy,  
where the unnormalized Bloch state $\ket{\psi_{\rm flat}}=(0,\,g,\,- f)^{T}$ has no weight on A orbitals. 
$f(\bk)$ and $g(\bk)$ are determined by the lattice geometry and hopping phases, and when they simultaneously vanish at a $\bk$ point, a band touching with other dispersive band(s) occurs as an immovable singularity~\cite{SinguFlatClassificationPRB2019,SinguFlatAdvPhysX2021}. 
We consider an example of such nearest-neighbor hopping SFB model on a dice lattice geometry~\cite{vidal1998aharonov,DicePRB1986},
\begin{eqnarray}
\hat{H}_0 =&& \sum_l E_A\,\hat{A}^{\dagger}_{l} \hat{A}_{l} \label{hamiltonian0} \\
&&- \sum_{\braket{l,m}}
\left( t\, e^{i \phi_1^{l,m}} \hat{A}^{\dagger}_l \hat{B}_m 
+ t\, e^{i \phi_2^{l,m}} \hat{A}^{\dagger}_l \hat{C}_m + h.c. \right), \notag
\end{eqnarray}
with a $2\pi/3$ flux threading each rhombus [Fig.~\ref{Fig1}(a)] which breaks time-reversal symmetry. This corresponds to the bipartite limit of the nearest-neighbor tight-binding model of twisted bilayer MoTe$_2$~\cite{HongyiNSR2020}, where A orbital detuning $E_A\ge0$, and the flux is from real-space Berry curvature in the moir\'e. The hopping between B and C orbitals has a small amplitude in twisted bilayer MoTe$_2$~\cite{xu2024maximally}, and is switched off here. 
One identifies
\begin{eqnarray}
        f(\bk) &=& -t\left( e^{-i \bs{k}\cdot \bs{d}_1} + e^{i 2\pi/3} e^{-i \bs{k}\cdot \bs{d}_2} + e^{-i 2\pi/3}  e^{-i \bs{k}\cdot \bs{d}_3}\right) \notag \\
        g(\bk) &=&  t\left(e^{ i \bs{k}\cdot \bs{d}_1} + e^{i 2\pi/3} e^{  i \bs{k}\cdot \bs{d}_2} + e^{-i 2\pi/3}  e^{ i \bs{k}\cdot \bs{d}_3}\right) ~\label{fg}
\end{eqnarray}
where $\bs{d}_{1,2,3}$ are nearest-neighbor vectors from A to B orbitals. $f$ and $g$ both vanish at the $\Gamma$ point, leading to a quadratic band touching between the flat and lower dispersive bands when $E_A>0$ [Fig.~\ref{Fig1}(b)], while a linear band touching emerges with all three bands intersecting at the $\Gamma$ point when $E_A=0$ [supplementary Fig.~\ref{Fig:LinearTouching}(a)].
The overcomplete set of zero-energy modes, including compact localized states and non-contractible loop states, is illustrated in supplementary Fig.~\ref{Fig:CLS}, which are the defining characters of a SFB~\cite{BandTouchingBalents2008,SinguFlatClassificationPRB2019}.

We note the crucial role of the bipartite symmetry of this model, which refers to the scenario where lattice sites can be divided into two groups with hopping allowed across groups but forbidden within each group of uniform onsite energy~\cite{DicePRB1986,FlatBandReview2018}.
In our case, one group consists of A orbitals and the other consists of B and C orbitals. From the perspective of real-space topology~\cite{BandTouchingBalents2008,SinguFlatClassificationPRB2019}, exact flatness and band touching of SFB is the consequence of over-completeness of zero-energy modes under the bipartite symmetry (Supplemental Material Sec.~S2). Therefore, the quadratic band touching under $E_A>0$ here is  qualitatively different from the ones discussed in Ref.~\cite{SunKaiQuadraticPRL2009}, where the quadratic touching can exist without exact band flatness and is protected by time-reversal and rotational symmetries.

Interestingly, the SFB here exhibits features of a Chern band. While ill-defined at the $\Gamma$ point, the Berry curvature $\Omega$ and quantum metric tensor $\Cal{G}$ are smooth and nearly uniform everywhere else [Figs.~\ref{Fig1}(d, e)]. The integral of $\Omega$ without the $\Gamma$ point yields $2\pi$ on the SFB, and $-2\pi$ on the lower dispersive band.
Upon a singular non-periodic gauge transformation $\ket{\psi'_{\rm flat}}=e^{-i\theta}\ket{\psi_{\rm flat}}$, where $\theta$ is the polar angle of $\bk$,
the normalized Bloch state $\ket{\psi'_{\rm flat}}$ is continuous, which approaches $(0,\,-i,\,i)^T/\sqrt{2}$ from all directions towards $\Gamma$. In this gauge, the berry connection $\bA'=i\braket{\psi'_{\rm flat}|\nabla \psi'_{\rm flat}}$ also becomes continuous and smooth around $\Gamma$ [Fig.~\ref{Fig1}(c)] (see Supplemental Material~\cite{supp}). 
We also note that the maximum quantum distance~\cite{SinguFlatNature2020} between Bloch states around the $\Gamma$ point is vanishing here, which also distinguishes this SFB from existing cases~\cite{SinguFlatAdvPhysX2021}.

\begin{figure}[t]
	\centering
	\includegraphics[width=3.4in]{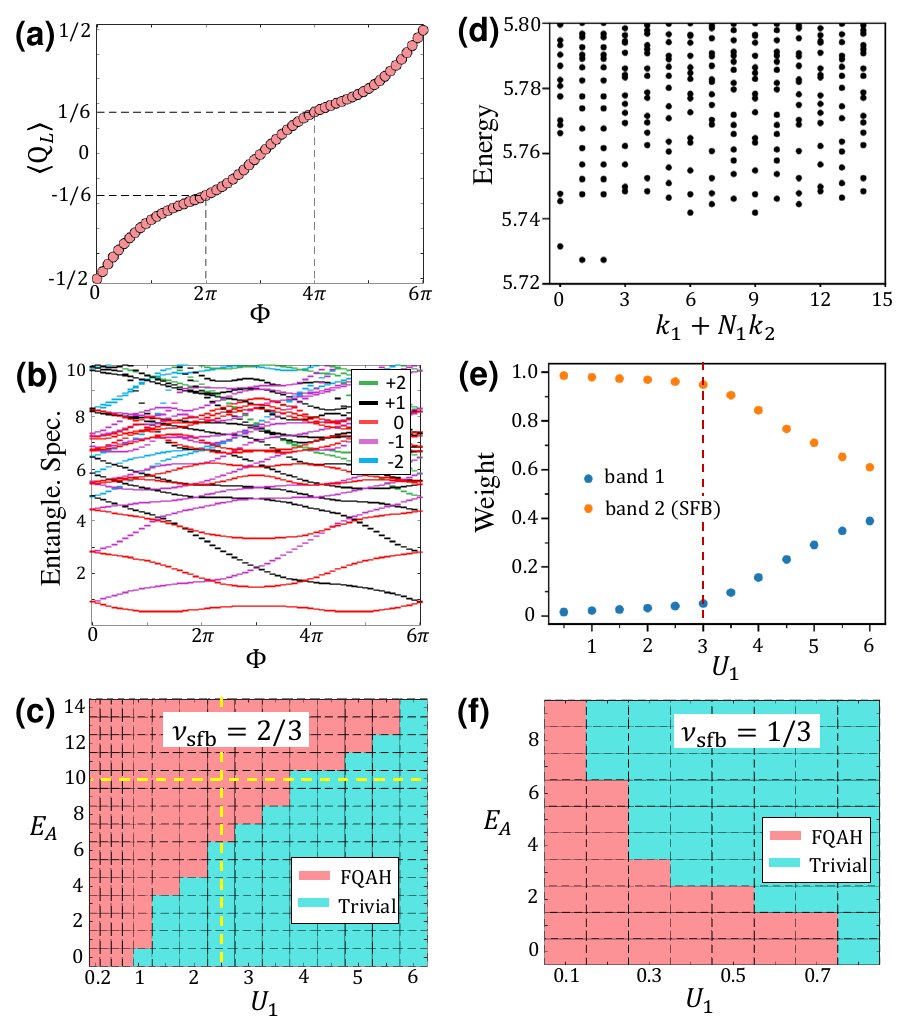}
	\caption{(a--c) Real-space DMRG calculations accounting all bands at $\nu_{\rm sfb} = 2/3$: FQAH effect demonstrated by charge pumping (a) and entanglement spectrum flow (b) upon flux insertion, at $E_A = 10$, $U_1 = 1.5$; (c) $E_A$--$U_1$ phase diagram, magenta color denotes the FQAH phase, and blue for a topologically trivial phase. (d--e) ED calculations on the two touching bands at $\nu_{\rm sfb} = 2/3$: (d) many-body eigenspectrum at $E_A = 10$, $U_1 = 0.5$; (e) band projections of the ground states at $E_A = 10$ (corresponding eigenspectra in supplementary Fig.~\ref{Fig:ED}). (f) Phase diagram at $\nu_{\rm sfb} = 1/3$ from DMRG.}
	\label{Fig2}
\end{figure}

{\bf FQAH effect at fractional SFB fillings} -- 
To properly account for the effect of band touching in the many-body correlation, we perform real-space DMRG calculations with the tight-binding Hamiltonian Eq.~(\ref{hamiltonian0}) and nearest-neighbor repulsion $H_{\rm int} =  \sum_{\braket{l,m}} U_1^{l,m} \hat{n}_l \hat{n}_m$ of spinless fermions, on a cylinder geometry 
(details in Supplemental Materials~\cite{supp}). 
We find FQAH phases at $1/3$ and $2/3$ fillings of the SFB in both the case of quadratic ($E_A\ne0$) and linear ($E_A=0$) band touching, which are signified by fractionally quantized charge pumping between open boundaries of the cylinder upon the adiabatic flux insertion. We focus on the quadratic touching case in the main text, and results for the linear touching case are shown in supplementary Figs.~\ref{Fig:LinearTouching}(b, c).

Figure~\ref{Fig2}(a) presents an example of the charge pumping at SFB filling factor $\nu_{\rm sfb}=2/3$ (the total filling factor $\nu = \nu_{\rm sfb} +1 = 5/3$). One charge quantum is pumped after three flux periods, signaling a fractionally quantized Hall conductivity of $e^2/(3h)$~\cite{NiuQianPRB1985}. 
The entanglement spectrum flow in Fig.~\ref{Fig2}(b) also provides evidence for a $1/3$ quantized charge pumping per cycle: the spectrum is shifted by one particle sector after $6\pi$ flux insertion.  
Figure~\ref{Fig2}(c) presents the phase diagram at $\nu_{\rm sfb}=2/3$ in the parameter space spanned by the interaction strength $U_1$ and onsite energy $E_A$. It consists of the FQAH phase in a wide parameter space characterized by the $1/3$ quantized charge pumping, and a trivial metallic phase of vanishing charge gap. 
The phase boundary is also corroborated by the discontinuity in entanglement spectrum~\cite{EsHaldanePRL,U1ES}, as shown in supplementary Fig.~\ref{Fig:ES} for the scans along the vertical and horizontal dashed lines in Fig.~\ref{Fig2}(c).

We also perform two-band ED calculations which provide complementary information that corroborates the DMRG findings. Figure~\ref{Fig2}(e) presents the band projection of the ground states at $E_A=10$ with various $U_1$, and the corresponding many-body eigenspectra are shown in Fig.~\ref{Fig2}(d) and supplementary Fig.~\ref{Fig:ED}. For $U_1 < 3$, the spectra show characteristics of FQAH phase, agreeing well with the DMRG phase diagram in Fig.~\ref{Fig2}(c). Interestingly, these FQAH ground states are predominantly weighted in the SFB, despite the band touching and strong interaction exceeding the spectral window spanned by the two touching bands. 
This is reminiscent of the findings in certain Chern insulator models where FQAH states exist and weight predominantly on the flat Chern band when interaction strength well exceeds the band gap~\cite{kourtis2024fractional,GrushinPRB2015}.
Upon transition to the metallic phase, the weight on the dispersive band starts to grow rapidly.

At $\nu_{\rm sfb} = 1/3$ (total filling $\nu = 4/3$), FQAH phase is also found, but in a narrow region where $U_1$ is small [Fig.~\ref{Fig2}(f)]. 
In this FQAH phase, two charge quanta are pumped upon $6\pi$ flux insertion (supplementary Fig.~\ref{Fig:fill43}), corresponding to a fractionally quantized Hall conductivity of $2e^2/(3h)$. 
At both $\nu_{\rm sfb} = 1/3$ and $2/3$, the lower boundary of FQAH phase as function of interaction strength $U_1$ is not observed within our computation capacity. The phase diagrams start from $U_1=0.1\sim0.2$, below which the small charge gap demands larger bond dimension for convergence.

{\bf Band touching and stability of FQAH} -- 
For both $\nu_{\rm sfb} = 1/3$ and $2/3$ FQAH phases, transition to a topologically trivial phase occurs upon the increase of interaction strength, but the $\nu_{\rm sfb} = 2/3$ FQAH phase is obviously more robust with a much larger upper critical $U_1$. To elucidate this sharp contrast, self-consistent Hartree-Fock calculations are performed at the closest integer fillings (supplementary Fig.~\ref{Figs_HF}), where the interaction-renormalized quasiparticle bands can in general provide some insight. 
$\nu_{\rm sfb} = 1/3$ corresponds to 1/3 particle added to the empty SFB at $\nu=1$.
At this integer filling, as $U_1$ is ramped up, the Hartree-Fock mean field leads to a small gap $\Delta_{\Gamma}$ at the quadratic band touching [supplementary Fig.~\ref{Figs_HF}(a)]~\cite{SunKaiQuadraticPRL2009}, whose contour line shows remarkable resemblance to the FQAH phase boundary at $\nu_{\rm sfb}=1/3$. 
We note the Fock term manifests as an effective hopping $t'_{\rm eff} (\hat{B}^{\dagger}_l \hat{C}_m + h.c.)$, with $t'_{\rm eff}$ proportional to $U_1$ and the nearest-neighbour inter-orbital coherence $\braket{C^{\dagger}_mB_l}$. The Hartree term generates an energy offset $\delta_{\rm eff}$, between the B and C orbitals, proportional to the inter-orbital polarization. Either term will break the bipartite symmetry, thereby gap the band touching and destroy exact band flatness [cf. Figs.~\ref{Fig3}(c, d)]. 

$\nu_{\rm sfb} = 2/3$ can be considered as 1/3 hole added on top of the full SFB filling at $\nu=2$. With B and C orbitals near fully filled, especially at large A orbital detuning, the inter-orbital coherence and polarization at $\nu=2$ have tiny magnitudes compared to the $\nu=1$ case, therefore opening a comparable $\Delta_{\Gamma}$ requires a much larger $U_1$. This is in line with the stability of $\nu_{\rm sfb} = 2/3$ FQAH phase against the increase of $U_1$.
The $\nu_{\rm sfb}=2/3$ FQAH state also has better thermal stability, characterized by a charge neutral excitation gap that can reach $\sim 0.03t$ while the charge gap can reach $\sim 0.1t$ (details in Supplemental Material Secs.~6 and 7).

\begin{figure}[t]
	\centering
	\includegraphics[width=3.4in]{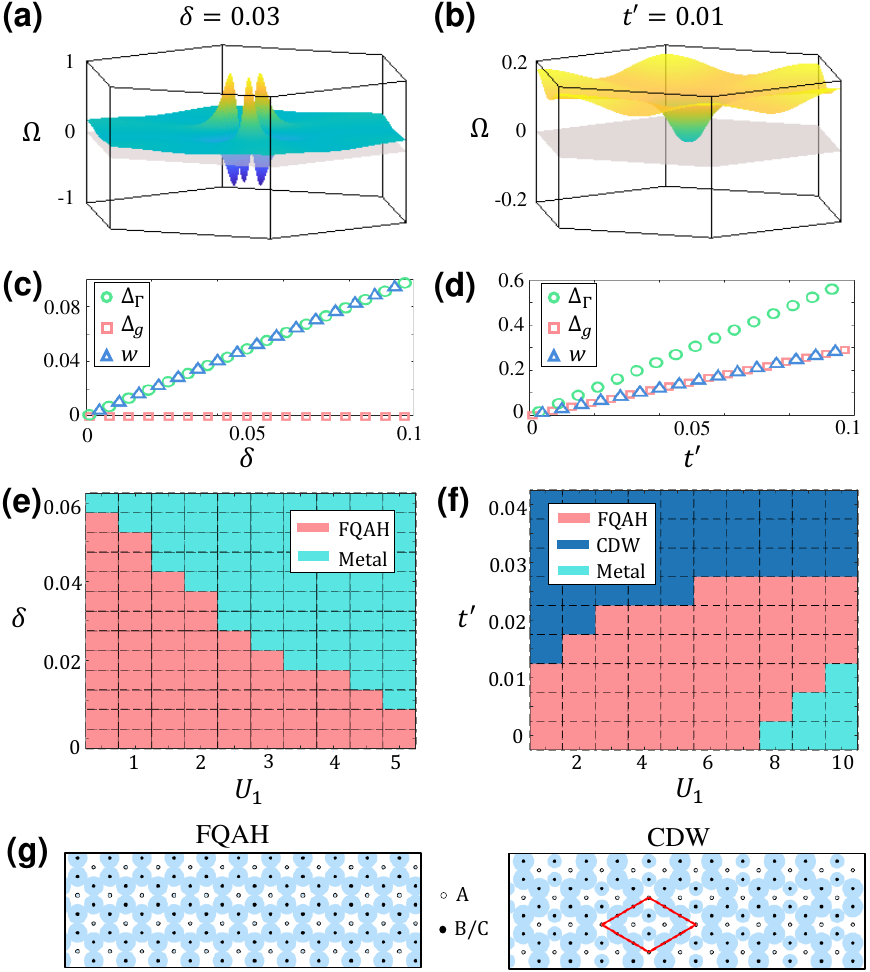}
	\caption{\textbf{Breaking of band touching and stability of FQAH effect.} (a, b) Berry curvature distribution in the perturbed flat band, when the band touching is gapped by a energy offset $\delta$ or small hopping $t'$ between B and C orbitals. (c, d) Bandgap ($\Delta_g$) between the two lower bands, their gap at $\Gamma$ point ($\Delta_\Gamma$), and bandwidth of the perturbed flat band ($w$), as functions of $\delta$ and $t'$. $E_A=8$ in panels (a--d). (e) $U_1$--$\delta$ phase diagram at $\nu_{\rm sfb}=2/3$. $t'=0$, $E_A=15$. (f) $U_1$--$t'$ phase diagram at $\nu_{\rm sfb}=2/3$. $\delta=0$, $E_A=15$. FQAH, metallic and CDW phases are denoted by magenta, light blue and dark blue colors respectively. (g) Examples of charge distributions of FQAH phase and CDW phase. Particle populations on each orbitals are denoted by the size of the blue circles. Red diamond denotes the $\sqrt{3}\times\sqrt{3}$ supercell in the CDW state. }
	\label{Fig3}
\end{figure}

Analysis of the ground state wave functions at the fractional fillings from DMRG calculations supports the observation that gapping of touching point in the interaction renormalized band accompanies the quench of the FQAH effect. In supplementary Fig.~\ref{Fig:ChargeGap}, the evolution of the many-body charge gap  at $\nu_{\rm sfb}=2/3$ as a function of $U_1$ is plotted along with the polarization between B and C orbitals.  
After a critical $U_1$ within the FQAH phase, weak orbital polarization and thus a finite $\delta_{\rm eff}$ through the Hartree term spontaneously emerges [supplementary Fig.~\ref{Fig:ChargeGap}(d)], coinciding with the decrease of the many-body charge gap. The transition from FQAH to the trivial metallic phase is accompanied by a further significant increase in the orbital polarization [supplementary Fig.~\ref{Fig:ChargeGap}(d)].

We note that gapping the band touching by either a small $t'_{\rm eff}$ or $\delta_{\rm eff}$ term turns the SFB into an isolated Chern band with nearly flat dispersion [supplementary Fig.~\ref{Figs_HF}(f)]. 
However, this comes at the cost of a strong inhomogeneity in the Berry curvature [cf. Figs.~\ref{Fig3}(a, b)]. With a larger gap, the inhomogeneity is spread out over a larger $k$-space region, which is unfavorable for FQAH state to form~\cite{WangJieCDWFCI,RoyPRB2014}.

The counter-intuitive role of band touching is further corroborated by examining phase diagram upon adding the B-C hopping $t'(\hat{B}^{\dagger}_l \hat{C}_m + h.c.)$ and energy offset $\delta$ to the single-particle Hamiltonian $\hat{H}_0$. 
Figure~\ref{Fig3}(e) shows the phase diagram in $U_1$--$\delta$ space, fixing $E_A=15$. 
A small $\delta$ to gap the touching point in the single-particle bands can significantly reduce the stability of FQAH phase against the increase of interaction strength. 
Figure~\ref{Fig3}(f) shows the phase diagram in $U_1$--$t'$ space at $E_A=15$. 
When $t'$ is small, a transition from FQAH to metallic phase at large $U_1$ is also observed, as a continuation of the $t'=0$ phase diagram.
In this regime $t'$ can push the FQAH to metal transition towards a larger $U_1$.
Increasing $t'$ slightly further destroys the FQAH phase and drives the system into a $\sqrt{3}\times\sqrt{3}$ charge density wave (CDW).
Figure~\ref{Fig3}(g) shows examples of the carrier distributions in FQAH and CDW phases respectively.

All these comparisons and analysis suggest that the FQAH states here are inherently associated with the SFB rather than an interaction renormalized isolated Chern band. The FQAH effect exists here in the weak $U_1$ limit where the Hartree-Fock bands fully retain the SFB feature, and in the moderate $U_1$ regime where the band touching is minimally gapped.
This finding is in stark contrast to existing paradigms, where opening of a sizable gap by interaction~\cite{guo2023theory,herzog2023moir,dong2023anomalous, dong2023theory} or single-particle effect~\cite{okamoto2022topological} to isolate a Chern band is favorable for the FQAH states. 


\begin{figure}[t]
	\centering
	\includegraphics[width=2.8in]{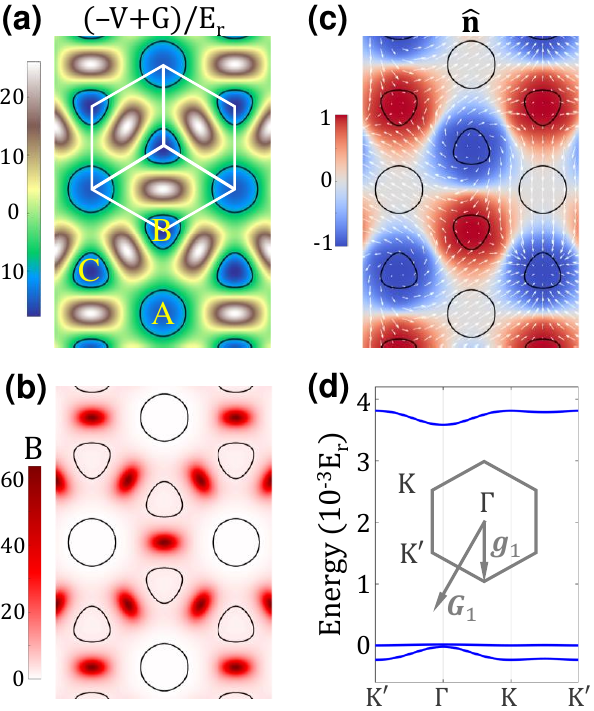}
	\caption{ \textbf{$2\pi$-flux dice lattice and SFB for ultracold atoms.} (a) Potential landscape in units of $E_r=\frac{\hbar^2}{2m}|G_1|^2$. The energy has been shifted 
    to place the barrier height between A and B/C sites at $0$. (b) Pseudo-magnetic field rescaled by unit cell area and flux quantum. The flux per unit cell is $2 \pi$. (c) Spatial texture of $\hat{\textbf{n}}$. Background color denotes $n_z$, and arrows for in-plane vector $(n_x,\,n_y)$. (d) Energy bands under $V_0\approx-19.75E_r$ and $V_1\approx-23.8E_r$. The energies have been shifted to place the flat band around zero energy.}
	\label{Fig:System}
\end{figure}

{\bf Realization of the SFB in optical lattice} -- 
We generalize here the optical flux lattice scheme initially proposed in Ref.~\cite{CooperPRL2011}. Consider atoms of two long-lived levels  
moving in optical ﬁelds: $
\hat{H}=\frac{\hat{p}^2}{2m}-\bs{V}(\bs{r}) \cdot \hat{\bs{\sigma}}$, 
where $\hat{\bs{\sigma}}$ is the vector of Pauli matrices. 
The off-diagonal $V_{x,y}$ are the interspecies optical coupling generated with three travelling waves~\cite{CooperPRL2011}:  $V_x=V_0\sum_{i=1}^{3}\cos(\bs{g}_{i}\cdot\bs{r})$, $V_y=-V_0\sum_{i=1}^{3}\sin(\bs{g}_{i}\cdot\bs{r})$.
$V_z$ corresponds to a species-dependent optical potential, which we consider a form generated by three standing waves: $V_z=-V_1\sum_{i=1}^{3}\sin(\bs{G}_{i}\cdot\bs{r})$. 
The wavevectors $\bs{G}_1=-(\frac{1}{\sqrt{3}},\,1) \frac{2\pi}{L}$ and 
$\bs{g}_1=-(0,\, \frac{2}{3}) \frac{2\pi}{L}$ [Fig.~\ref{Fig:System}(d)], 
and $\bs{G}_{2,3}$ and $\bs{g}_{2,3}$ are their $120^\circ$ degree rotations respectively. The magnitude and direction of $\bs{V}$ will be denoted by $V(\bs{r})$ and $\hat{\textbf{n}}(\bs{r})$.

Adiabatic motion on the local ground states with Bloch vectors along $\hat{\textbf{n}}(\bs{r})$ is then governed by an effective Hamiltonian: $\hat{H}_{\text{adb}}=\frac{1}{2m}[\hat{\bs{p}}+e\bs{A}(\bs{r})]^2-V(\bs{r}) + G(\bs{r})$
where atoms experience a periodic scalar potential with geometric correction $G=\frac{\hbar^2}{8m}(\nabla \hat{\textbf{n}})^2$~\cite{AdySternPRL1992}, and a pseudo-magnetic field $\bs{B}=\nabla\times\bs{A}=\frac{\hbar}{2e}\hat{\textbf{n}}\cdot(\partial_{x}\hat{\textbf{n}}\times\partial_{y}\hat{\textbf{n}}) \textbf{z}$. 
The magnetic flux threading a unit cell equals $2\pi$, as determined by the solid angle enclosed by $\hat{\textbf{n}}$. For the example in Fig.~\ref{Fig:System}, the three local minima in the potential landscape define the orbitals, where tunneling between the degenerate B and C sites is suppressed by a significant barrier [Fig.~\ref{Fig:System}(a)], realizing a $2\pi$--flux dice lattice. The lowest three energy bands [Fig.~\ref{Fig:System}(d)] indeed reproduce all features of the non-interacting tight-binding Hamiltonian featuring the SFB. 
This cold atom platform offers a clean and tunable arena to examine the counter-intuitive role of band touching in stabilizing FQAH in the SFB.

{\bf Acknowledgment} -- We thank Bin-Bin Chen, Wei Zheng, Zhen Zheng, Jie Wang and Qian Niu for helpful discussions. Wenqi Yang thanks Tao Xiang, Kang Wang and Runze Chi for guidance on learning DMRG. The work is supported by Research Grant Council of Hong Kong SAR (HKU SRFS2122-7S05, AoE/P-701/20, A-HKU705/21), and New Cornerstone Science Foundation.

\bibliography{FluxedDiceRefs}



\clearpage 

\renewcommand{\thefigure}{S\arabic{figure}}
\renewcommand{\thesection}{S\arabic{section}}
\renewcommand{\thepage}{S\arabic{page}}
\renewcommand{\theequation}{S\arabic{equation}}
\setcounter{figure}{0}
\setcounter{section}{0}
\setcounter{page}{1}
\setcounter{equation}{0}

\title{Supplemental Material for ``Fractional quantum anomalous Hall effect in a singular flat band''}
\maketitle
\onecolumngrid
\tableofcontents
\newpage
\section{Generic 2-orbital SFB model with quadratic band touching and its connection to the 3-orbital model}

Without loss of generality, we will set the flat band at zero energy in the following. The minimal singular flat band (SFB) scenario corresponds to a two-orbital model with a quadratic band touching. In the following we discuss the generic properties of such a minimal model and explain its connection to the three-orbital model introduced in the main text.

\textbf{Generic two-orbital model with zero-energy flat band.} A $2\times2$ Hamiltonian with unnormalized eigenstates $\ket{\psi_{1,2}}$ and eigenenergies $\varepsilon_{1,2}(\bk)$ can generally be expressed as $H=\varepsilon_1(\bk) P_1 + \varepsilon_2(\bk) P_2$,
where $P_i=\ket{\psi_{i}}\bra{\psi_{i}}/\braket{\psi_i|\psi_i}$ is the projection operator satisfying $P_1+P_2=1$. If $\ket{\psi_2}$ is chosen as a flat-band eigenstate with zero energy ($\varepsilon_2\equiv 0$), the Hamiltonian reads $H=\varepsilon_1(\bk)(1-P_2)$.
Therefore, a two-orbital Hamiltonian with flat-band eigenstate of the form $\ket{\psi_{2}}=(g(\bk),\,-f(\bk))^T$ at the zero energy can be generically expressed as $H_{\rm 2O}=\lambda_{\bk}\begin{pmatrix} |f|^2&f^{*}g\\g^{*}f&|g|^2 \end{pmatrix}$, where $\lambda_{\bk}$ is an arbitrary real function with Brillouin zone periodicity~\cite{SinguFlatDesignGrafPRB2021}. To simplify notations, in the following we define $R=(f,\,g)$, thus $H_{\rm 2O}=\lambda_{\bk}R^{\dagger}R$. 

\textbf{Properties of the two-orbital model.} The band structure of $H_{\rm 2O}$ is composed of $\varepsilon_1(\bk)=\lambda_{\bk}RR^{\dagger}=\lambda_{\bk}(|f|^2+|g|^2)$ 
and flat band $\varepsilon_2(\bk)=0$, and the associated normalized eigenvectors are $\ket{\psi_1}=(f^{*},\,g^{*})^{T}/\sqrt{RR^{\dagger}}$
and $\ket{\psi_2}=(g,\,-f)^{T}/\sqrt{RR^{\dagger}}$. Although $H_{\rm 2O}\ket{\psi_2}=0$ is generally satisfied independent of the choices of $\lambda_{\bk}$, $f$ and $g$, the flat band is not necessarily singular with a touching point. A SFB can be achieved by setting e.g. $\lambda_{\bk}$ to a constant, and choosing $f$ and $g$ properly such that they vanish simultaneously at the same wave vector, i.e., $f(\bk_0)=g(\bk_0)=0$, leading to a touching point at $\bk_0$. In lattice models, $f$ and $g$ depend on the hopping parameters and Bloch phases [see e.g., Eq.~(\ref{fg}) of the main text].  We note that destructive interferences of the Bloch phases leading to $f(\bk_0)=g(\bk_0)=0$ can occur even in the absence of time-reversal and spatial symmetries, but the presence of symmetries could help to stabilize the touching point.

\textbf{Limitations of the two-orbital model.} It should be noted that the corresponding real-space lattice models of $H_{\rm 2O}$ usually are complicated, which involve hopping processes beyond the nearest neighbors. Moreover, it is desirable to have convenient tunability in the model to explore various correlated phases, which is however lacking in $H_{\rm 2O}$ as this would necessarily require changing hopping parameters and/or lattice geometries. These issues not only impose computational constraints on the numerical exploration of many-body phenomena, but also complicate the experimental implementation of the lattice model.

\textbf{Connection to the three-orbital model.} To avoid the above limitations, the three-orbital Hamiltonian $H_0=\begin{pmatrix}E_A&R\\R^{\dagger}&O\end{pmatrix}$ was introduced in Eq.~(\ref{eq:H0_k_space}) of the main text, where $O$ is a $2\times2$ null matrix and here $E_A>0$ is assumed. Importantly, $H_0$ reproduces the same SFB properties of $H_{\rm 2O}$, and moreover, it exhibits extra advantageous features as elaborated below. The three energy bands of $H_0$ in ascending order are $\varepsilon_1(\bk)=(E_A - \sqrt{E_A^2 + 4RR^{\dagger}})/2$, $\varepsilon_2(\bk)=0$ and $\varepsilon_3(\bk)=(E_A + \sqrt{E_A^2 + 4RR^{\dagger}})/2$, where $\varepsilon_1$ and the flat band $\varepsilon_2$ have touching point at the same $\bk_0$ as that of $H_{\rm 2O}$. The eigenvector of the flat band is identical to that of $H_{\rm 2O}$ up to an extra null component, i.e., $\ket{\Psi_2}=(0,\,g,\,-f)^{T}/\sqrt{RR^{\dagger}}$. The connection between $H_0$ and $H_{\rm 2O}$ becomes more transparent in the limit of $E_A\gg |f|,\,|g|$ or at low energy around the band touching. In these regimes, $H_0$ reduces to two separate sectors, i.e., $H_u=RR^{\dagger}/E_A+E_A$ characterizing the upper band, and $H_l=-R^{\dagger}R/E_A$ accounting for the flat band touching with the lower dispersive band. $H_l$ is identical to $H_{\rm 2O}$ with $\lambda_{\bk}=-E_A^{-1}$.

\textbf{Advantages of the three-orbital model.} As shown in the main text, $H_0$ can be realized on a lattice with nearest-neighbor hopping only, which reduces computational cost and facilitates experimental implementation of the model. It also exhibits several advantageous features without introducing formidable complexities in numerics: (i) $E_A$ serves as a convenient tuning parameter for achieving different correlated phases via its effects on the dispersive bands and their eigenstates; (ii) SFBs with different band touching configurations, i.e., quadratic or linear, can be easily realized by setting $E_A\ne0$ or $E_A=0$; (iii) Different perturbations that break the band touching [e.g., Fig.~\ref{Fig3}(c, d) of the main text] can be readily introduced to examine their effects on correlated behaviors; (iv) With $f$ and $g$ in Eq.~(\ref{fg}) of the main text, $H_0$ reproduces the model of twisted bilayer MoTe$_2$ by turning on the hopping between B and C orbitals~\cite{HongyiNSR2020}, thus the results can potentially shed light on the correlated phenomena in moir\'e materials; (v) By varying $f$ and $g$ properly, SFB models with different topological and quantum geometric properties (e.g., Berry phase different from $\pm2\pi$ around the touching point) can be constructed.

Finally, we remark that there exist different perspectives for understanding the emergence of the flat band in $H_0$. In quantum optics, $H_0=\begin{pmatrix}E_A&R\\R^{\dagger}&O\end{pmatrix}$
with $R=(f,\,g)$ shares the same structure as the so-called $\Lambda$-scheme having two degenerate lower levels without direct coupling [Fig.~\ref{Fig1}(a) inset of main text], where the resultant dark state is the counterpart of the flat-band eigenstate. More generally, with $R$ of a generic $\Cal{M}\times\Cal{N}$ matrix, the mismatch in the dimension of the null space (i.e., space of zero-energy modes) of $R$ and $R^{\dagger}$, which equals $\Cal{N}-\Cal{M}$, serves as a `topological' index determining the number of flat bands in $H_0$. This index is widely used in topological mechanics~\cite{SUSYTopoMechNatPhys2014}, bipartite lattices~\cite{DicePRB1986,LiebTheorem}, etc. Note that such perspectives do not distinguish the singular or non-singular nature of the bands (with or without band touching), which requires the detailed knowledge of $R$.

\section{Compact localized states and non-contractible loop states of the 2$\pi$-flux dice lattice}

The SFB host localized eigenstates-- the compact localized states (CLS), which are strictly bounded within a finite region~\cite{DicePRB1986}. In a lattice of $N$ unit cells, a total number of $N$ CLS centered in each cell can be constructed. 
Under periodic boundary condition, in a \textit{non-singular} flat band, these $N$ CLS form a complete basis; in contrast, the $N$ CLS are not linearly independent in a \textit{singular} flat band, thus additional states with the flat-band energy are needed to span the flat band completely~\cite{BandTouchingBalents2008,SinguFlatClassificationPRB2019}. These additional states are the non-contractible loop states (NLS), which are extended in one direction but compactly localized in the other. The NLS thus can be considered as defining characters of a SFB. The addition of NLS to the CLS makes the total number of independent states with the flat-band energy more than $N$, thus explaining the presence of band touching~\cite{BandTouchingBalents2008}. 

\begin{figure}[h]
	\centering
	\includegraphics[width=3.5in]{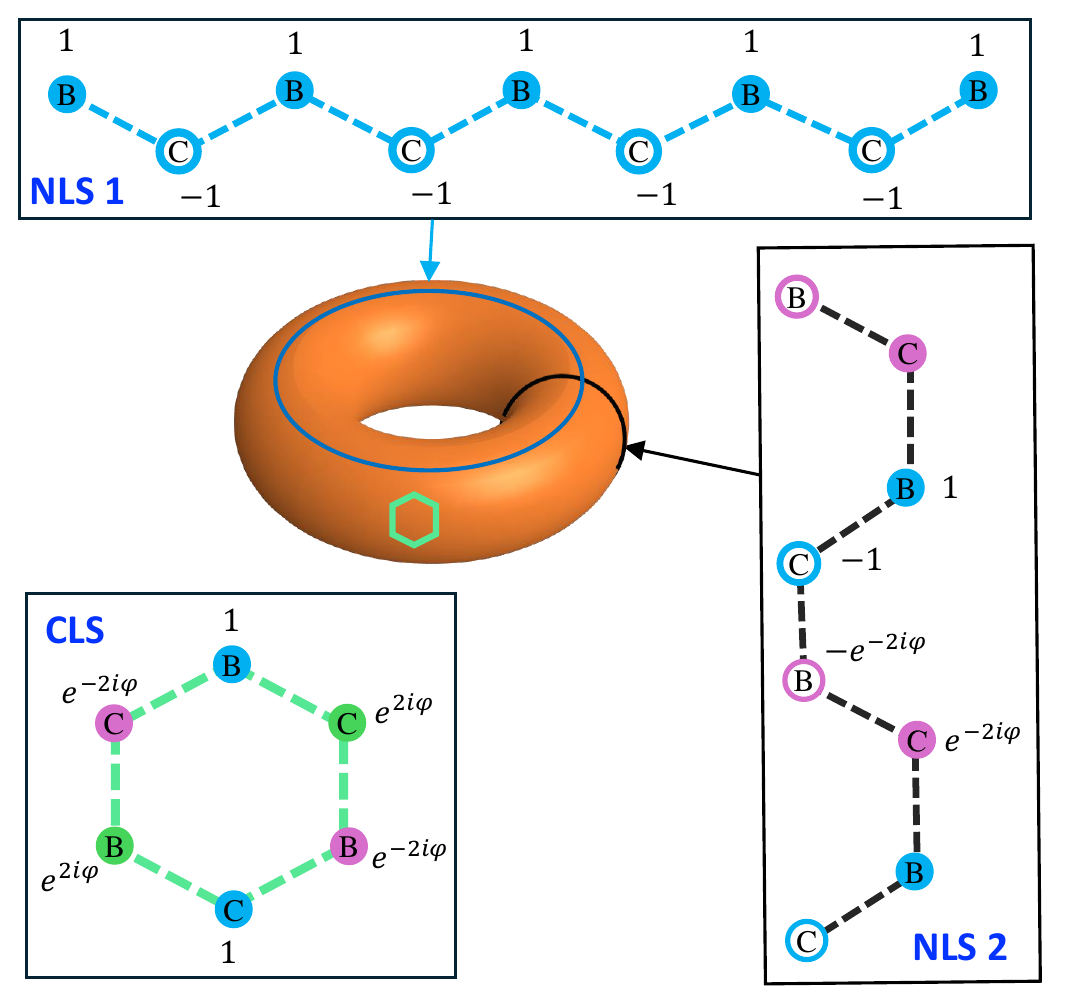}
	\caption{Schematics of the compact localized state (CLS) and two non-contractible loop states (NLS) of the 2$\pi$-flux dice lattice with periodic boundary condition (a torus), the existence of which dictates that the flat band with a touching point here is a singular flat band (SFB). Circled B and C dots denote the orbitals, the numbers represent their contributing weights to the CLS or NLS.}
    \label{Fig:CLS}
\end{figure}

For the 2$\pi$-flux dice lattice, $N-1$ independent CLS an be constructed on the hexagon formed by B and C orbitals, as shown schematically in Fig.~\ref{Fig:CLS} lower left panel. The numbers by the orbitals represent the amplitude of the CLS at the sites. Meanwhile, there exist two independent NLS that are also independent of the CLS, as schematically shown in Fig.~\ref{Fig:CLS} top and right panels. A total number of $N+1$ independent zero-energy modes ($N-1$ CLS and 2 NLS) is consistent with the presence of one touching point between the flat band and the lower dispersive band [Fig.~1(b) of the main text]. 

\section{Remarks on non-interacting properties around the $\Gamma$ point}

For small $k$ around $\Gamma$, one finds that $\ket{\psi_{\rm flat}}$ approaches $\ket{\psi_0}=e^{i\theta}(0,\,-i,\,i)^T/\sqrt{2}$, where $\theta$ is the polar angle of $\bk$. 
We point out that $\ket{\psi_0}$ cannot fully capture the quantum geometric properties of the system around $\Gamma$, e.g., the Berry curvature, thus higher order terms in the Bloch state are important here.
Let us consider the Berry curvature explicitly. The Berry connection associated with $\ket{\psi_0}$ is $\bs{A_0}=i\braket{\psi_0|\nabla \psi_0}=-\nabla\theta=-\hat{\theta}/k$, where $\hat{\theta}=(-k_y,\,k_x)/k$ is the unit vector in the polar direction. The resultant Berry curvature is $\bs{\Omega}_{0}=\nabla\times\bs{A_0}=-2\pi \delta(\bk)\hat{\bs{z}}$, which vanishes except at the $\Gamma$ point. 
By comparing to the nonzero results of $\Omega$ in Fig.~1(d) of the main text, it is clear that $\bs{\Omega}_{0}$ resulting from $\ket{\psi_0}$ is insufficient to properly account for the geometric properties around the $\Gamma$ point, thus higher order terms are crucial for the quantum geometry. 

Nevertheless, the above results suggest that if one performs a singular and non-periodic gauge transformation $\ket{\psi'_{\rm flat}}=e^{-i\theta}\ket{\psi_{\rm flat}}$, the Berry conncetion becomes $\bA'=\bA-\bA_0\approx(-k_y,\,k_x)/16$ around the $\Gamma$ point, where $\bA$ and $\bA'$ are the Berry connection from $\ket{\psi_{\rm flat}}$ and $\ket{\psi'_{\rm flat}}$, respectively. $\bA'$ yields a constant Berry curvature equals $1/8$ around the $\Gamma$ point, which is consistent with the numerical results in Fig.~1(d) of the main text.

Here we also provide the full expression of $\bA=(A_x\,A_y)$ in the gauge of $\ket{\psi_{\rm flat}}$. It reads
\begin{equation}
    \begin{aligned}
        A_x&=-\frac{1}{N^2}\left[\frac{1}{2}\sin(\bk\cdot\bs{b}_1)+\frac{1}{2}\sin(\bk\cdot\bs{b}_2)-\sin(\bk\cdot\bs{b}_3)\right]\\
        A_y&=-\frac{1}{N^2}\left[\frac{\sqrt{3}}{2}\sin(\bk\cdot\bs{b}_1)-\frac{\sqrt{3}}{2}\sin(\bk\cdot\bs{b}_2)\right]
    \end{aligned},
\end{equation}
where $N^2=|f|^2+|g|^2=6-2\sum_{i=1}^{3}\cos(\bk\cdot\bs{b}_i)$, $\bs{b}_1=\bs{d}_1-\bs{d}_2$, $\bs{b}_2=\bs{d}_3-\bs{d}_1$, and $\bs{b}_3=\bs{d}_2-\bs{d}_3$. One notices that both $N^2$ and the terms in the square brackets become zero at the $\Gamma$ point, where there is a singularity. By keeping the linear-order contribution of $\bk$ in each term of $A_x$ and $A_y$, one finds $\bA\approx\bA_0$; while retaining higher orders of $\bk$ can also reproduce $\bA\approx\bA_0+\bA'$, which are consistent with results in the previous paragraphs.

Specifically, using Taylor expansion of $\cos x=1-x^2/2!+x^4/4!-x^6/6!+\cdots$, one finds
\begin{equation}
    \begin{aligned}
        N^2&=2\sum_{i=1}^{3}\left[\frac{(\bk\cdot\bs{b}_i)^2}{2!}-\frac{(\bk\cdot\bs{b}_i)^4}{4!}+\frac{(\bk\cdot\bs{b}_i)^6}{6!}-\cdots\right]
        =k^2\left[\sum_{i=1}^{3}\cos^2\alpha_i-2k^2\sum_{i=1}^{3}\frac{\cos^4\alpha_i}{4!}+2k^4\sum_{i=1}^{3}\frac{\cos^6\alpha_i}{6!}-\cdots\right]\\
        &=\frac{3}{2}k^2\left[1-\frac{4}{3}k^2\sum_{i=1}^{3}\frac{\cos^4\alpha_i}{4!}+\frac{4}{3}k^4\sum_{i=1}^{3}\frac{\cos^6\alpha_i}{6!}-\cdots\right]
    \end{aligned},
\end{equation}
where $\alpha_i$ is the angle between $\bk$ and $\bs{b}_i$ satisfying $\cos\alpha_1=\cos(\theta+\pi/6)$, $\cos\alpha_2=-\cos(\theta-\pi/6)$, and $\cos\alpha_3=\sin\theta$, and we have used $\sum_{i=1}^{3}\cos^2\alpha_i=3/2$. For $k\approx0$, one can approximate
\begin{equation}
    \frac{1}{N^2}\approx\frac{2}{3k^2}\left[1+\left(\frac{4}{3}k^2\sum_{i=1}^{3}\frac{\cos^4\alpha_i}{4!}-\frac{4}{3}k^4\sum_{i=1}^{3}\frac{\cos^6\alpha_i}{6!}+\cdots\right)+\left(\frac{4}{3}k^2\sum_{i=1}^{3}\frac{\cos^4\alpha_i}{4!}-\frac{4}{3}k^4\sum_{i=1}^{3}\frac{\cos^6\alpha_i}{6!}+\cdots\right)^2+\cdots\right].
\end{equation}
While using Taylor expansion of $\sin x=x-x^3/3!+x^5/5!-\cdots$, one can rewrite the terms in the square brackets of $A_x$ and $A_y$ as
\begin{equation}
    \begin{aligned}
        -\frac{1}{2}\sin(\bk\cdot\bs{b}_1)-\frac{1}{2}\sin(\bk\cdot\bs{b}_2)+\sin(\bk\cdot\bs{b}_3)&=k\left(-\frac{1}{2}\cos\alpha_1-\frac{1}{2}\cos\alpha_2+\cos\alpha_3\right)\\
        &-k^3\left(-\frac{1}{2}\frac{\cos^3\alpha_1}{3!}-\frac{1}{2}\frac{\cos^3\alpha_2}{3!}+\frac{\cos^3\alpha_3}{3!}\right)\\
        &+k^5\left(-\frac{1}{2}\frac{\cos^5\alpha_1}{5!}-\frac{1}{2}\frac{\cos^5\alpha_2}{5!}+\frac{\cos^5\alpha_3}{5!}\right)\\
        &-\cdots\\
        -\frac{\sqrt{3}}{2}\sin(\bk\cdot\bs{b}_1)+\frac{\sqrt{3}}{2}\sin(\bk\cdot\bs{b}_2)&=k\left(-\frac{\sqrt{3}}{2}\cos\alpha_1+\frac{\sqrt{3}}{2}\cos\alpha_2\right)\\
        &-k^3\left(-\frac{\sqrt{3}}{2}\frac{\cos^3\alpha_1}{3!}+\frac{\sqrt{3}}{2}\frac{\cos^3\alpha_2}{3!}\right)\\
        &+k^5\left(-\frac{\sqrt{3}}{2}\frac{\cos^5\alpha_1}{5!}+\frac{\sqrt{3}}{2}\frac{\cos^5\alpha_2}{5!}\right)\\
        &-\cdots
    \end{aligned}.
\end{equation}
One notices that the linear-order contributions in Eq.~(S4) (i.e., $\frac{3}{2}k\sin\theta$ and $-\frac{3}{2}\cos\theta$) combined with the first term in $1/N^2$ (i.e., $\frac{2}{3k^2}$) leads to $\bA_0$. While all the other terms in $A_{x,\,y}$ are analytic, especially, the contributions linear in $k$ lead to $\bA'$.

\section{Details of DMRG calculations}

DMRG are performed for spinless fermions 
with the many-body tight-binding Hamiltonian $\hat{H} = \hat{H}_0 + \sum_{\braket{l,m}} U_1^{l,m} \hat{c}^{\dagger}_m \hat{c}_m  \hat{c}^{\dagger}_l \hat{c}_l$ on a lattice of the cylinder geometry with open boundary condition in the $x$ direction and periodic boundary condition in the $y$ direction. A flux $\Phi$ threading the cylinder is introduced through the twisted boundary condition~\cite{NiuQianPRB1985}. The total number of lattice sites is $N = N_y\times N_x\times3$, where $N_y = 4$ and $N_x = 24$ are the number of unit cells in the $y$ and $x$ directions. We have checked convergence by confirming that larger values of bond dimension $D$ give identical results. Calculations at integer filling used bond dimensions up to $D = 256$. At $\nu_{\rm sfb} = 2/3$ filling, a maximal bond dimension of $D = 300$ is used. At $\nu_{\rm sfb} = 1/3$ filling, the maximal $D = 500$ due to the smaller charge gap. The DMRG simulation is performed using the ITensor library with $U(1)$ symmetry~\cite{ITensor-r0.3}. 

In the calculations, nearest-neighbor repulsion between B and C orbitals is set slightly different ($0.9U_1$) from their repulsion with A orbital ($U_1$), to reflect the different nature of these orbitals, whereas the choice only have minor quantitative effect on the phase boundaries.



\section{Entanglement spectrum scans across the phase boundary at $\nu_{\rm sfb} = 2/3$}

\begin{figure}[h]
	\centering
	\includegraphics[width=4.5in]{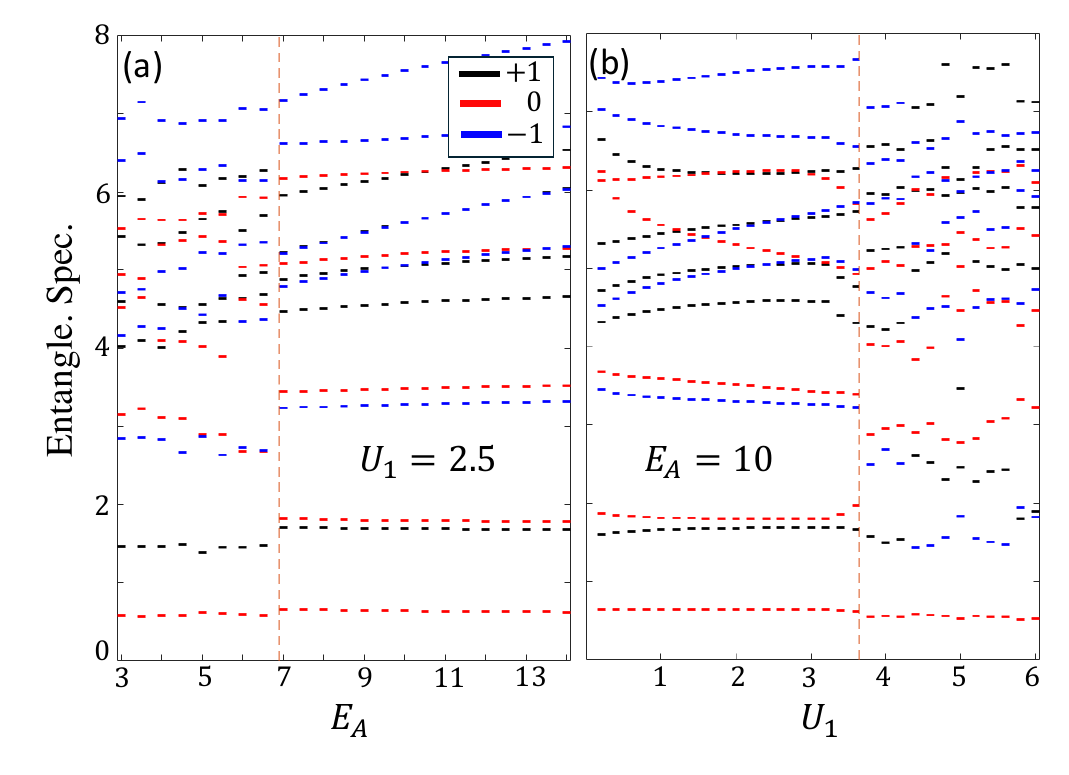}
	\caption{(a, b) Entanglement spectrum as a function of $E_A$ and $U_1$ at $\nu_{\rm sfb} = 2/3$. Different charge sectors are distinguished by different colors~\cite{U1ES}. The transition points are marked by the dashed vertical lines where sharp discontinuities of entanglement spectrum can be observed, signaling the transition between the FQAH phase and the metallic phase.}
	\label{Fig:ES}
\end{figure}

The entanglement spectrum, $\epsilon_\alpha$, is obtained using the relation $\epsilon_\alpha = -ln(\Lambda_\alpha^2)$~\cite{EsHaldanePRL}, where $\Lambda _\alpha$ denote Schmidt values obtained from the Schmidt decomposition at the bond connecting left and right halves of the lattice. As shown in Fig.~\ref{Fig:ES}, for both entanglement spectrum scans marked by the yellow dashed lines in Fig.~\ref{Fig2}(c) of the main text, there is a sharp discontinuity at the phase boundary, signaling the transition between the FQAH phase and the metallic phase.

\section{Evolution of the many-body charge gap and orbital polarization at $\nu_{\rm sfb} = 2/3$}

\begin{figure}[ht]
	\centering
	\includegraphics[width=5.3in]{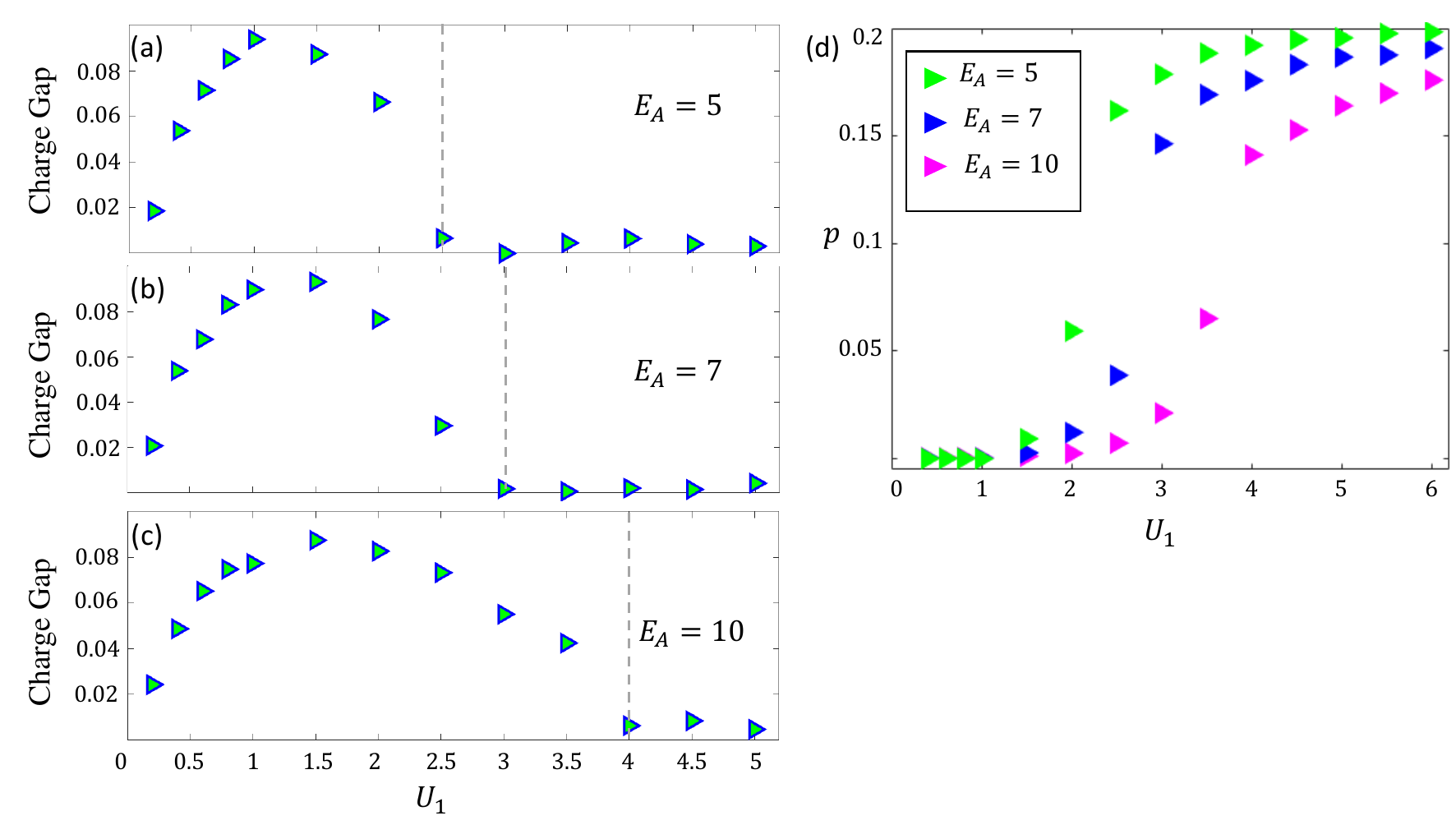}
	\caption{(a--c) Charge gap as a function of $U_1$ for different $E_A$ at $\nu_{\rm sfb} = 2/3$. The dashed vertical lines mark the boundary separating the FQAH and metallic phases. (d) Evolution of orbital polarization, $p = (\rho_B-\rho_C)/(\rho_B+\rho_C)$, as a function of $U_1$ at $\nu_{\rm sfb} = 2/3$.}
	\label{Fig:ChargeGap}
\end{figure}

The particle-hole excitation gap (charge gap) of the system is given by: 
\begin{eqnarray}
\Delta_c &=& \mu^p - \mu^h.
\end{eqnarray}
where $\mu^p$ ($-\mu^h$) is the energy of adding a particle (hole) to the system~\cite{ChargegapPRB}:
\begin{equation}
\begin{aligned}
\mu^p &= E^p - E_0 \\
-\mu^h &= E^h - E_0
\end{aligned},
\end{equation}
$E_0$ here is the ground state energy without the particle and hole, and $E^p$ ($E^h$) is the total energy of the state with an additional particle (hole), both of which are computed using DMRG calculations.

As shown in Figs.~\ref{Fig:ChargeGap}(a)--(c), FQAH phase has a finite charge gap which first increases then decreases as a function of $U_1$, reaching zero at the phase boundary and remaining vanishing in the metallic phase. The initial rise in the charge gap can be attributed to the stronger interaction. After a critical $U_1$ within the FQAH phase, weak orbital polarization $p$ spontaneously emerges [Fig.~\ref{Fig:ChargeGap}(d), also see Fig.~\ref{Figs_polarization}(b) in a larger parameter space]. A finite $p$ breaks the band touching and introduces inhomogeneity to the quantum geometry of the second band. As a result, the charge gap of the FQAH state decreases. The transition from FQAH to the metallic phase occurs upon a significant increase in $p$.


\section{Charge neutral excitation gap at $\nu_{\rm sfb} = 2/3$}\label{Sec:SuppNeutralGap}

The charge neutral excitation gap ($\Delta_{\rm cn}$) is calculated using DMRG, which can shed light on the thermal stability of the FQAH states~\cite{HongyuThermal}. We first get the ground state wave function $\ket{\Psi_0}$ in the matrix product state (MPS) form and the ground state energy $E_0$ of the many-body Hamiltonian $\hat{H}$ via DMRG simulation. $U(1)$ symmetry is used in our calculation, and the total particle number of $\ket{\Psi_0}$ is set at the value corresponding to the filling factor $\nu_{\rm sfb}=2/3$. We then apply DMRG simulation on a new matrix product operator $\hat{H} + w \ket{\Psi_0}\bra{\Psi_0}$ to get another MPS wave function $\ket{\Psi_1}$, where $\ket{\Psi_0}$ acts as the penalty state and $w$ is a positive constant which needs to be large enough in order to minimize the overlap between $\ket{\Psi_1}$ and $\ket{\Psi_0}$. Since the $U(1)$ symmetry is implemented, $\ket{\Psi_1}$ is the lowest-energy eigenstate of $\hat{H}$ among states within the same particle-number sector as $\ket{\Psi_0}$ while orthogonal to $\ket{\Psi_0}$. The expectation value $E_1=\bra{\Psi_1}\hat{H}_0\ket{\Psi_1}$ is the energy of the particle-conserved first excited state, and the energy difference $E_1-E_0=\Delta_{\rm cn}$.

\begin{figure}[t]
	\centering
	\includegraphics[width=3in]{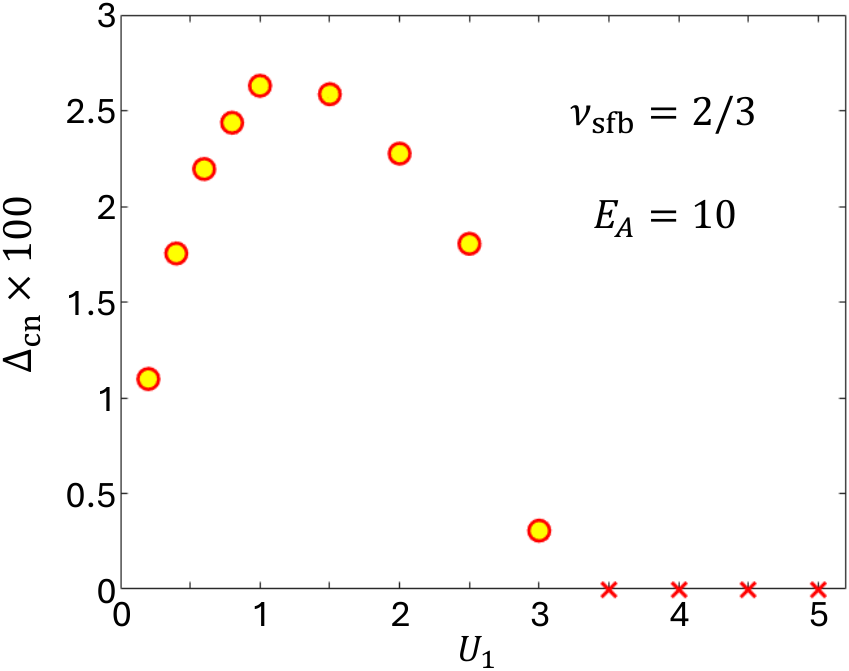}
	\caption{Charge neutral excitation gap $\Delta_{\rm cn}$ as a function of $U_1$ for $E_A=10$ and $\nu_{\rm sfb}=2/3$. $\Delta_{\rm cn}$ becomes negligible when $U_1\gtrsim3.5t$ and are marked by red crosses.}
	\label{Fig:NeutralGap}
\end{figure}

Figure~\ref{Fig:NeutralGap} shows the evolution of $\Delta_{\rm cn}$ at $\nu_{\rm sfb}=2/3$ and $E_A=10$ as a function of $U_1$. The trend is generally the same as that of the charge gap [cf. Fig.~\ref{Fig:ChargeGap}(c)], i.e., it first increases then decreases with $U_1$ in the FQAH regime. 
For $U_1\gtrsim3.5t$, $\Delta_{\rm cn}$ becomes negligible within our computation capacity [denoted by red crosses in Fig.~\ref{Fig:NeutralGap}]. In the FQAH phase, $\Delta_{\rm cn}$ is $\sim 0.01-0.03$ in units of the hopping integral $t$, and is $\sim 1/3$ of the charge gap.
We note that the singular flat band here has a rather uniform Berry curvature distribution favorable for thermal stability~\cite{JainCurvatureGap}.


\section{Charge pumping simulations at $\nu_{\rm sfb} = 1/3$}

\begin{figure}[h]
	\centering
	\includegraphics[width=2.5in]{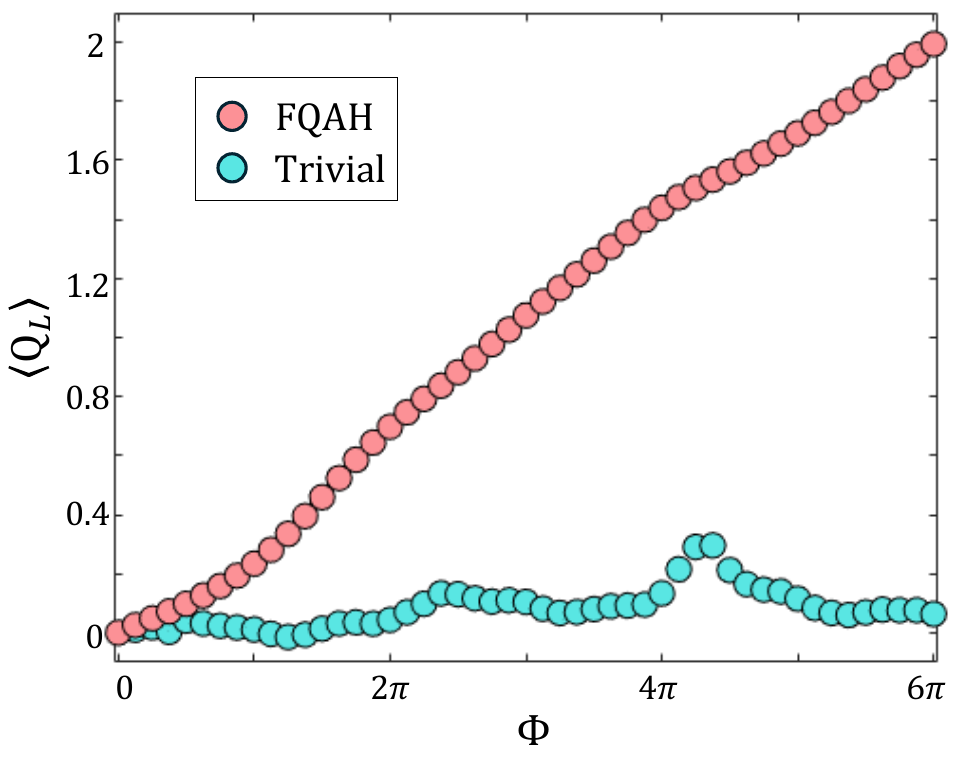}
	\caption{Charge pumping simulation results of the FQAH state ($U_1 = 0.1$, $E_A = 7$) and topologically trivial state ($U_1 = 0.5$, $E_A = 8$) at $\nu_{\rm sfb} = 1/3$.}
	\label{Fig:fill43}
\end{figure}

The $\nu_{\rm sfb} = 1/3$ phase diagram in the $U_1-E_A$ parameter space is shown in Fig.~\ref{Fig2}(f). 
FQAH phase only exists in a narrow region where $U_1$ is small. 
Fig.~\ref{Fig:fill43} shows examples of the charge pumping simulations of the FQAH phase and topologically trivial phase, respectively.

\section{Gap opening at the touching point by interaction and FQAH stability}

\textbf{Hartree-Fock} -- The Hartree-Fock results at integer fillings $\nu=1$ and $\nu=2$ can shed light on the different behaviors at $\nu_{\rm sfb}=1/3$ and $\nu_{\rm sfb}=2/3$ respectively. We first discuss the details for the self-consistent Hartree-Fock calculations. The single-particle Hamiltonian $\hat{H}_0(\bm{k})$ is defined in the main text. While the interaction Hamiltonian is
\begin{equation}
\hat{H}_{I}=\sum_{\braket{i,j}} U_1^{A,B} A^{\dagger}_{i}A_{i}B^{\dagger}_jB_j+ U_1^{A,C} A^{\dagger}_{i}A_{i}C^{\dagger}_jC_j+ U_1^{B,C} C^{\dagger}_{i}C_{i}B^{\dagger}_jB_j
\end{equation}
where we set $U_1^{A,C}=U_1^{A,B}=U_1$ and $U_1^{B,C}=0.9U_1$ as in DMRG calculations. The summation $\braket{i,j}$ is over all nearest-neighbour pairs. Transforming the interaction Hamiltonian into momentum space yields
\begin{equation}
\hat{H}_{I}=\sum_{\bm{k}_1-\bm{k}_4 \in BZ,\bm{\delta}_m}\frac{U_1}{N}A^{\dagger}_{\bm{k}_1}B^{\dagger}_{\bm{k}_2}B_{\bm{k}_4}A_{\bm{k}_3}\delta_{\lceil \bm{k}_1+\bm{k}_2-\bm{k}_3-\bm{k}_4\rceil ,0}e^{i(\bm{k}_4-\bm{k}_2)\cdot \bm{\delta}_m}+\cdots
\end{equation}
where $\cdots$ denotes similar terms involving other nearest-neighbour interaction, $N$ is the number of unit cells, and we have defined the operator $\lceil\cdots\rceil$ to send its argument back to the Brillouin Zone (BZ), $\bs{\delta}_m$ is the vector pointing from unit cell where $i$ orbitals live to unit cell where $j$ orbitals live and the $m$ subscript denotes different neighbours.  The Hartree-Fock approximation is done by replacing the
four-fermion operators with all its possible contractions.

\begin{equation}
\begin{split}
A^{\dagger}_{\bm{k}_1}B^{\dagger}_{\bm{k}_2}B_{\bm{k}_4}A_{\bm{k}_3}\to& \braket{A^{\dagger}_{\bm{k}_1}A_{\bm{k}_3}}B^{\dagger}_{\bm{k}_2}B_{\bm{k}_4}\delta_{\bm{k}_1,\bm{k}_3}\delta_{\bm{k}_2,\bm{k}_4}+\braket{B^{\dagger}_{\bm{k}_2}B_{\bm{k}_4}}A^{\dagger}_{\bm{k}_1}A_{\bm{k}_3}\delta_{\bm{k}_1,\bm{k}_3}\delta_{\bm{k}_2,\bm{k}_4}\\
&-\braket{A^{\dagger}_{\bm{k}_1}B_{\bm{k}_4}}B^{\dagger}_{\bm{k}_2}A_{\bm{k}_3}\delta_{\bm{k}_1,\bm{k}_4}\delta_{\bm{k}_2,\bm{k}_3}-\braket{B^{\dagger}_{\bm{k}_2}A_{\bm{k}_3}}A^{\dagger}_{\bm{k}_1}B_{\bm{k}_4}\delta_{\bm{k}_1,\bm{k}_4}\delta_{\bm{k}_2,\bm{k}_3}
\end{split}
\end{equation}
where $\braket{\cdots}$ is evaluated with respect to the Slater Determinant ground state to be defined later. We define the projector $\rho(\bm{k})$ as
\begin{equation}
\rho_{\alpha,\beta}(\bm{k})\equiv \braket{\beta^{\dagger}_{\bm{k}}\alpha_{\bm{k}}}\quad\quad(\alpha,\beta=A,B,C)
\end{equation}
Then the interacting Hamiltonian is reduced to the mean-field form
\begin{equation}
\begin{split}
\hat{H}_{I,MF}(\bm{k})=&\sum_{\bm{k}_1}\frac{3U_1}{N}\rho(\bm{k}_1)_{A,A}B^{\dagger}_{\bm{k}}B_{\bm{k}}+\sum_{\bm{k}_1}\frac{3U_1}{N}\rho(\bm{k}_1)_{B,B}A^{\dagger}_{\bm{k}}A_{\bm{k}}\\
&-\sum_{\bm{k}_1,\bm{\delta}_m}\frac{U_1}{N}e^{i(\bm{k}_1-\bm{k})\cdot \bm{\delta}_m}\rho(\bm{k}_1)_{B,A}B^{\dagger}_{\bm{k}}A_{\bm{k}}-\sum_{\bm{k}_1,\bm{\delta}_m}\frac{U_1}{N}e^{-i(\bm{k}_1-\bm{k})\cdot \bm{\delta}_m}\rho(\bm{k}_1)_{A,B}A^{\dagger}_{\bm{k}}B_{\bm{k}}+\cdots
\end{split}
\end{equation}

$\hat{H}_0+\hat{H}_I$ is diagonalized in the $A,B,C$ orbital basis, with wavefunction coefficients $u^n_{\alpha}(\bm{k})$, where $n=1,2,3$ labels the band. The Hartree-Fock orbitals $d^{\dagger}_{\bm{k},\bm{n}}$ are expressed in the original orbitals as
\begin{equation}
d^{\dagger}_{\bm{k},n}=\sum_{\alpha}u^n_{\alpha}(\bm{k})\alpha^{\dagger}_{\bm{k}}
\end{equation}

The Slater Determinant ground state is constructed from the vacuum $\ket{\Omega}$ as
\begin{equation}
\ket{\text{Slater}}=\prod_{n,\bm{k}\in \text{filled}}d^{\dagger}_{\bm{k},n}\ket{\Omega}
\end{equation}

Then, the projector is calculated from the wavefunction coefficients as
\begin{equation}
\rho_{\alpha,\beta}(\bm{k})=\sum_{n\in \text{filled}}u^n_{\alpha}(\bm{k})u^{n*}_{\beta}(\bm{k})
\end{equation}

This is a self-consistent process, since the projector and the mean-field Hamiltonian are dependent on each other. We use a random initial projector to locate the state with lowest energy $E$, which can be calculated as
\begin{equation}
E\equiv\braket{\text{Slater}|\hat{H}_0+\hat{H}_I|\text{Slater}}=\sum_{\bm{k}}\text{Tr}\left[\rho(\bm{k})\left(\hat{H}_0(\bm{k})+\frac{\hat{H}_{I,MF}(\bm{k})}{2}\right)\right]
\end{equation}

All our calculations are done on a $24\times24$ momentum space grid, we take the set of $\bm{k}$ to be
\begin{equation}
\bm{k}=\frac{N_1}{24}\bm{g}_1+\frac{N_2}{24}\bm{g}_1,\quad N_i=0,\cdots,23
\end{equation}
where $\bm{g}_i$ are primitive reciprocal lattice vectors.

\begin{figure}[t]
	\centering
	\includegraphics[width=7.2in]{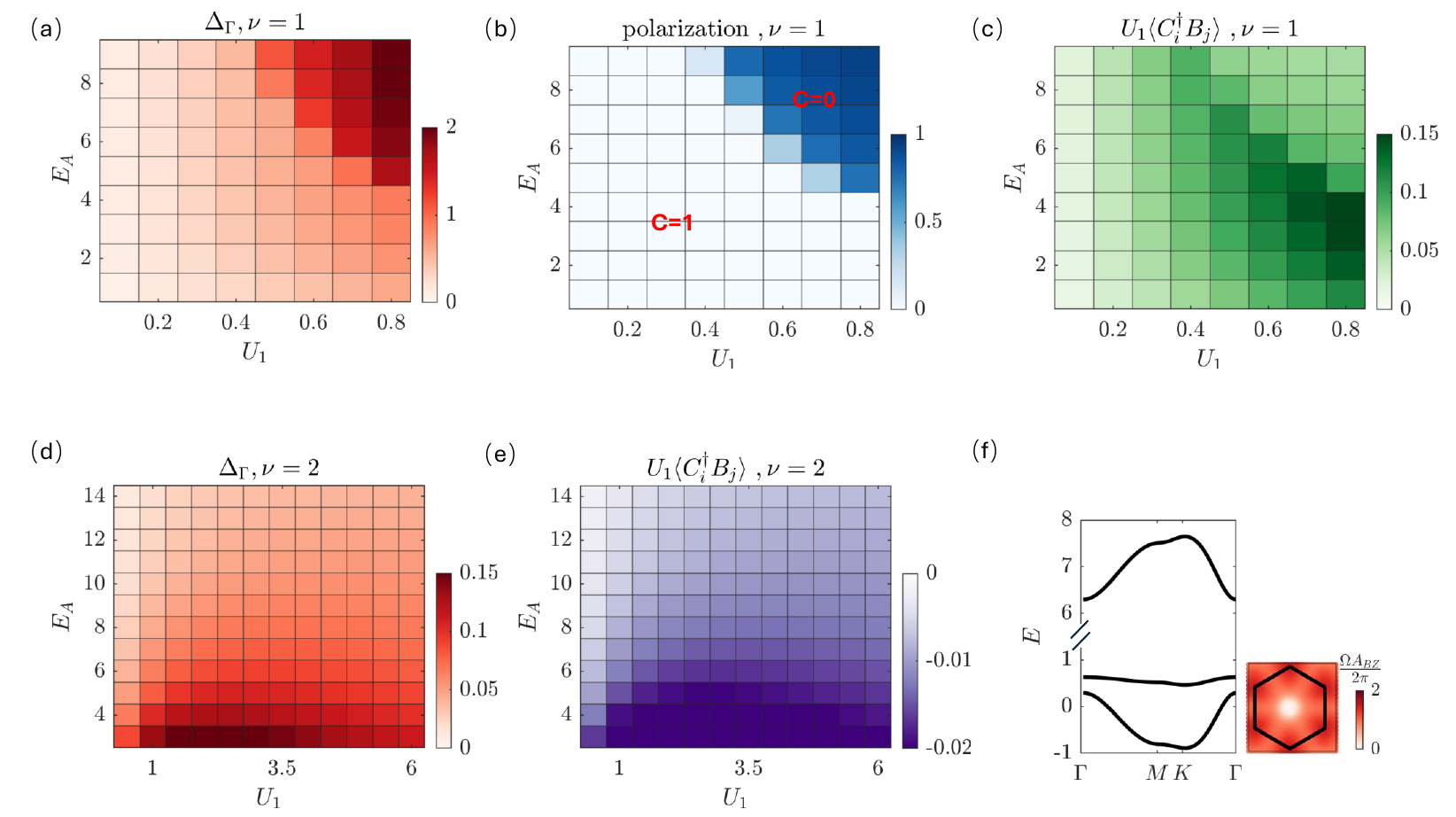}
	\caption{(a--e) Hartree-Fock Phase Diagram. (a) $\Delta_{\Gamma}$, the gap opening of at the $\Gamma$ point. (b) Orbital polarization $p = (\rho_B-\rho_C)/(\rho_B+\rho_C)$. (c) Nearest-neighbour inter-orbital coherence $\braket{C^{\dagger}_iB_j}U_1$. $C$ is the Chern number of the second quasiparticle band. (a--c) are done at filling factor $\nu=1$. (d, e) Similar results at $\nu=2$. The orbital polarization is vanishing, thus not shown. (f) Hartree-Fock band structure at $\nu=1$, $U_1=0.3$, $E_A=5.5$, corresponding to a point around the phase boundary in the regime where the ground state is topologically trivial at $\nu_{\rm sfb}=1/3$ in Fig.~2(f) of the main text. The right panel shows the Berry curvature of the second quasiparticle band.}~\label{Figs_HF}
\end{figure}

\begin{figure}[t]
	\centering
	\includegraphics[width=7.2in]{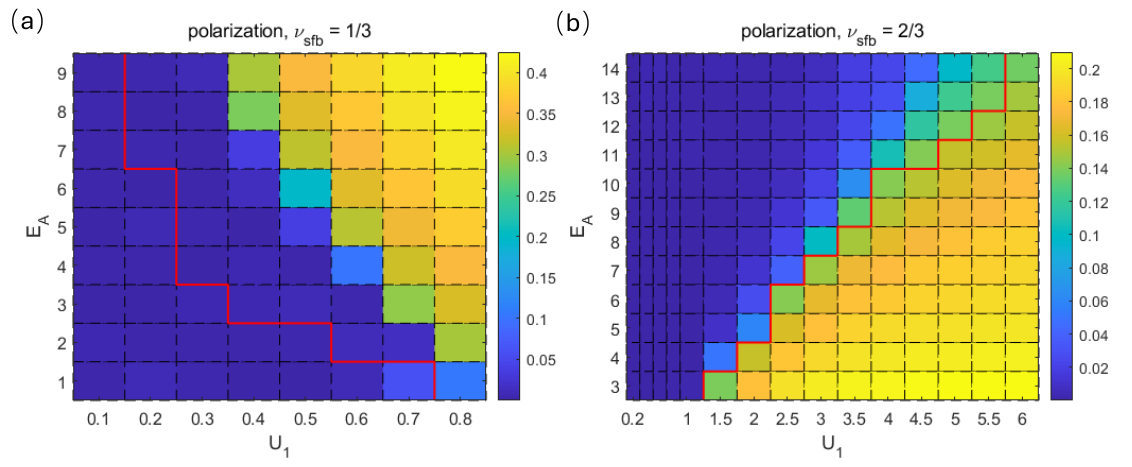}
	\caption{DMRG calculation results of polarization between B and C orbitals at fractional filling of (a) $\nu_{\rm sfb}=1/3$ and (b) $\nu_{\rm sfb}=2/3$. The phase boundary separating the FQAH and topologically trivial phases [see Figs.~2(c) and (f) of the main text] is marked by the red lines.}~\label{Figs_polarization}
\end{figure}

Fig.~\ref{Figs_HF} presents Hartree-Fock calculation results at total filling factor $\nu=1$ and $\nu=2$. As $U_1$ is ramped up, the interaction term leads to a small gap at the band touching point via the effective $t'$ from inter-orbital coherence, i.e., $t'_{\rm eff}=U_1\braket{C_i^{\dagger}B_j}$, and the effective $\delta$ from spontaneous orbital polarization, i.e., $\delta_{\rm eff}=U_1\left(\braket{n_B}-\braket{n_C}\right)$. 

For $\nu=1$, the touching point is gapped out by orbital coherence when $E_A$ and/or $U_1$ is small, which turns the SFB into a nearly flat Chern band with Chern number $C=1$. With larger $U_1$ and $E_A$, pronounced spontaneous orbital polarization between B and C orbitals emerges, and a large $\delta_{\rm eff}$ can turn the all quasiparticle bands into topologically trivial ones, which is reminiscent of the interaction-driven phase transition from integer QAH to site nematic phase discovered for the checkerboard and Kagome models in Ref.~\cite{SunKaiQuadraticPRL2009}. The contour line of $\Delta_{\Gamma}$, as shown in Fig.~\ref{Figs_HF}(a), shows remarkable resemblance to the FQAH phase boundary at $\nu_{\rm sfb}=1/3$ [Fig.~2(f) of the main text]. Notice that the FQAH state can only exist in the parameter range where $\Delta_{\Gamma}$ is small, indicating that gapping out of the touching point will quench the FQAH effect. Fig.~\ref{Figs_HF}(f) presents a typical quasiparticle band structure at $\nu=1$, where the touching point is gapped out such that the SFB becomes a well isolated Chern band. We note that, at this $U_1$ and $E_A$ parameter, the $\nu_{\rm sfb} = 1/3$ ground state is a topologically trivial one.

For $\nu=2$, $\Delta_{\Gamma}$ is negligible compared to that of the $\nu=1$ case [Figs.~\ref{Figs_HF}(d) vs (a)], thus the band touching is nearly intact. This is because B and C orbitals are near fully filled, the orbital polarization is totally suppressed and the orbital coherence is also very weak [Fig.~\ref{Figs_HF}(e)]. The robustness of the quasiparticle band touching point at $\nu = 2$ underlies the much larger FQAH phase region at $\nu_{\rm sfb} = 2/3$ than $\nu_{\rm sfb} = 1/3$ [compare Figs.~2(c) and (f) of the main text].

We remark that the Hartree-Fock approximation can reveal the variation of $\Delta_{\Gamma}$ as function of $U_1$ and $E_A$ at integer fillings,
while it tends to overestimate the energy gap. Therefore, the gap opening at the touching point could be smaller than the results shown in Figs.~\ref{Figs_HF}(a) and (d).

\textbf{DMRG} -- In Fig.~\ref{Figs_polarization}, we present DMRG results of orbital polarization in the $E_A$--$U_1$ parameter space at $\nu_{\rm sfb} = 1/3$ and $\nu_{\rm sfb} = 2/3$, respectively. The red lines mark the boundary separating the FQAH and topologically trivial phases [see Figs.~2(c) and (f) of the main text]. 

For the $\nu_{\rm sfb} = 2/3$ case [Fig.~\ref{Figs_polarization}(b)], the orbital polarization exhibits a sharp increase around the diagonal of the diagram, whose position agrees well with the boundary between the FQAH and topologically trivial phase [red curve, see Fig.~2(c) of the main text]. We have checked that the orbital coherence at $\nu_{\rm sfb} = 2/3$ is small and remains positive in the whole parameter space. A small positive orbital coherence and the obital polarization break the the touching point and turn the SFB into a Chern band with $C=1$, which is consistent with the value of Hall conductivity from charge pumping simulations.

For $\nu_{\rm sfb} = 1/3$ [Fig.~\ref{Figs_polarization}(a)], the phase boundary (red curve) lies deeper in the region where the orbital polarization is vanishing. This implies that, between the red curve and the light blue blocks (beyond which orbital polarization becomes significant), 
the relatively large orbital coherence as revealed in the Hartree-Fock bands at $\nu=1$ [Fig.~\ref{Figs_HF}(c)] is responsible for the quench of the FQAH state. 

\section{Details of exact diagonalization calculations and many-body eigenspectra}
We discuss the details of exact diagonalization calculation in this section. As addressed in the previous section, the interaction Hamiltonian in the momentum space reads

\begin{equation}
\hat{H}_{I}=\sum_{\bm{k}_1-\bm{k}_4 \in BZ,\bm{\delta}_m}\frac{U_1}{N}A^{\dagger}_{\bm{k}_1}B^{\dagger}_{\bm{k}_2}B_{\bm{k}_4}A_{\bm{k}_3}\delta_{\lceil \bm{k}_1+\bm{k}_2-\bm{k}_3-\bm{k}_4\rceil ,0}e^{i(\bm{k}_4-\bm{k}_2)\cdot \bm{\delta}_m}+\cdots,
\end{equation}
which can be expressed in terms of the single-particle band basis,
\begin{equation}
\hat{H}_{I}=\sum_{\substack{\bm{k}_1-\bm{k}_4 \in BZ,\bm{\delta}_m\\ n_1,n_2,n_3,n_4=1..3}}\frac{U_1}{N}\gamma^{\dagger}_{\bm{k}_1,n_1}\gamma^{\dagger}_{\bm{k}_2,n_2}\gamma_{\bm{k}_4,n_4}\gamma_{\bm{k}_3,n_3}\delta_{\lceil \bm{k}_1+\bm{k}_2-\bm{k}_3-\bm{k}_4\rceil ,0}e^{i(\bm{k}_4-\bm{k}_2)\cdot \bm{\delta}_m}U^*_{A,n_1}U^*_{B,n_2} U_{B,n_4}U_{A,n_3}+\cdots
\end{equation}
Here $\gamma^{\dagger}_{\bm{k},n}$ and $\gamma_{\bm{k},n}$ are respectively the creation and annihilation operators for single-particle bands, and are related to the original orbital operators by a $3 \times 3$ unitary matrix $U$,
\begin{equation}
\gamma^{\dagger}_{\bm{k,n}}=\sum_{\alpha}U_{\alpha n}(\bm{k})\alpha^{\dagger}_{\bm{k}},\quad \alpha^{\dagger}_{\bm{k}}=\sum_n U^*_{\alpha n} \gamma^{\dagger}_{\bm{k,n}}.
\end{equation}

\begin{figure}[t]
	\centering
	\includegraphics[width=6.5in]{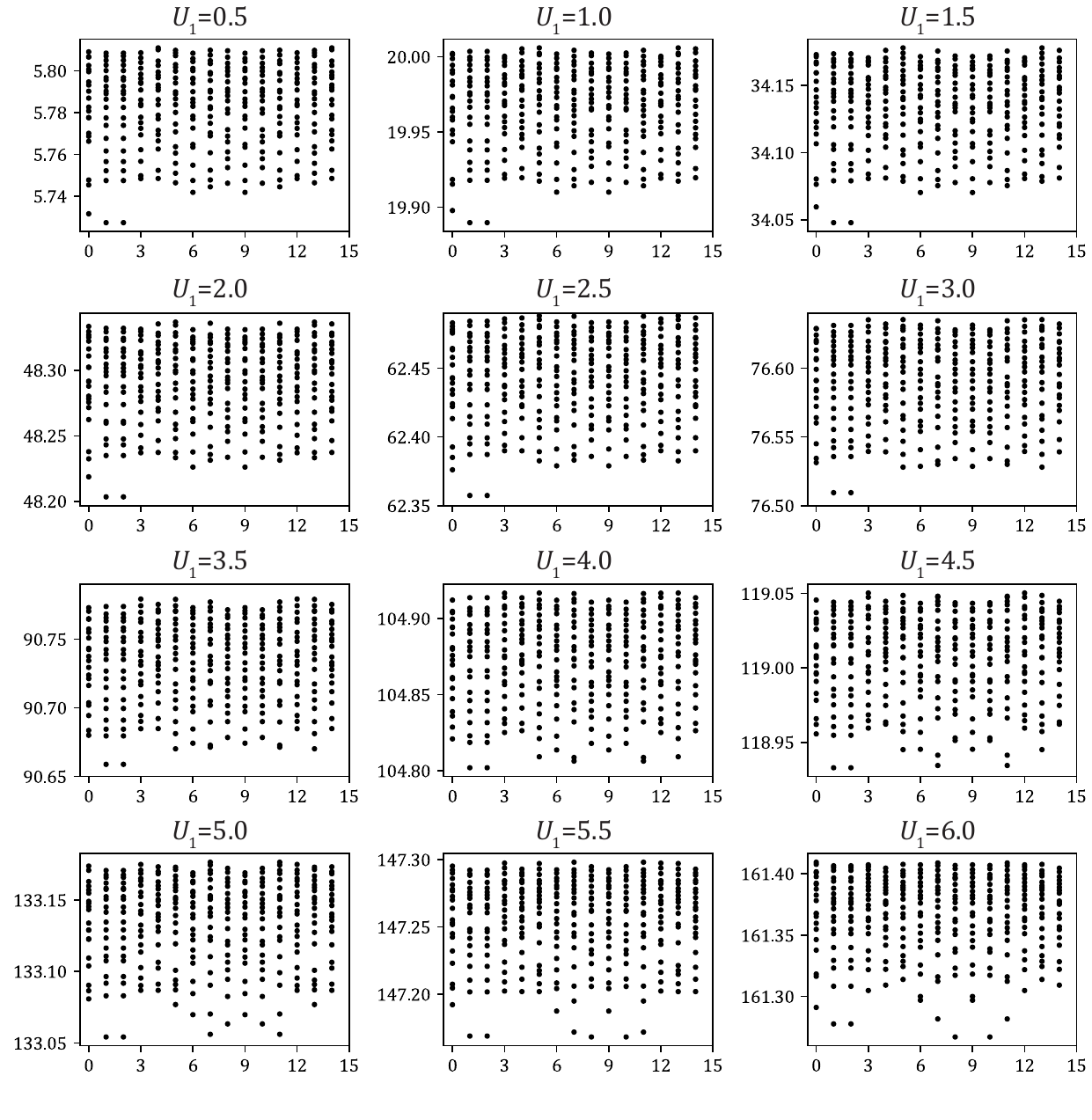}
	\caption{Many-body spectra under different interaction strength $U_1$, with an orbital detuning at $E_A=10$. The y-axis represents energy while the x-axis represents $k_1+N_1 k_2$ with $N_1=3$.}
	\label{Fig:ED}
\end{figure}

The calculations shown in the main text are done by projecting the single-particle Hamiltonian and the interaction  Hamiltonian to the lowest two single-particle bands, specifically,
\begin{equation}
\hat{H}_{0,p}=\sum_{n=1,2,\bm{k}\in BZ} E_n(\bm{k}) \gamma^{\dagger}_{\bm{k},n} \gamma_{\bm{k},n},
\end{equation}
and
\begin{equation}
\hat{H}_{I,p}=\sum_{\substack{\bm{k}_1-\bm{k}_4 \in BZ,\bm{\delta}_m\\ n_1,n_2,n_3,n_4=1,2}}\frac{U_1}{N}\gamma^{\dagger}_{\bm{k}_1,n_1}\gamma^{\dagger}_{\bm{k}_2,n_2}\gamma_{\bm{k}_4,n_4}\gamma_{\bm{k}_3,n_3}\delta_{\lceil \bm{k}_1+\bm{k}_2-\bm{k}_3-\bm{k}_4\rceil ,0}e^{i(\bm{k}_4-\bm{k}_2)\cdot \bm{\delta}_m}U^*_{A,n_1}U^*_{B,n_2} U_{B,n_4}U_{A,n_3}+\cdots
\end{equation}

The variational Hilbert space is taken to be spanned by all possible $N_p$-particle Slater determinant states constructed from the lowest two bands $\prod^{N_p}_{i=1}\gamma^{\dagger}_{\bm{k_i},n_i}\ket{\Omega}$. For this problem, momentum and particle number are good quantum numbers. Assume that the calculated ground state is $\ket{\eta}$, the band-occupation matrix is defined as

\begin{equation}
O_{n_1,n_2}(\bm{k})\equiv \braket{\eta|\gamma^{\dagger}_{\bm{k},n_1}\gamma_{\bm{k},n_2}|\eta},
\end{equation}
then the number of particle in band $i$ is $\mathcal{N}_i=\sum_{\bm{k}}O_{ii}(\bm{k})$. The weight presented in Fig.~2(e) of the main text is defined as $W_i\equiv 1-\frac{\mathcal{N}_i}{\mathcal{N}_1+\mathcal{N}_2}$. Finally, we study this problem on a finite-size torus. For a $3\times5$ system, as presented in Fig.~2(d) of the main text, the corresponding momentum space grid is
\begin{equation}
\bm{k}=\frac{k_1}{3}\bm{g}_1+\frac{k_2}{5}\bm{g}_2, ~~k_1=0,\cdots,2, \text{ and } k_2=0,\cdots,4.
\end{equation}
Fig.~\ref{Fig:ED} shows the extended data from exact diagonalization at $\nu_{\rm sfb}=2/3$ ($N=25$). For $U_1< 3.0$, there are three degenerate ground states at momentum sector $k_1+N_1k_2=0,1,2$, identical to that of the Laughlin state on a torus. The correct ground state degeneracy and correct momentum sector are strong evidence for FQAH in the system.


\section{Results of charge pumping in the case of a linear band touching at $E_A=0$}

\begin{figure}[h]
	\centering
	\includegraphics[width=6.5in]{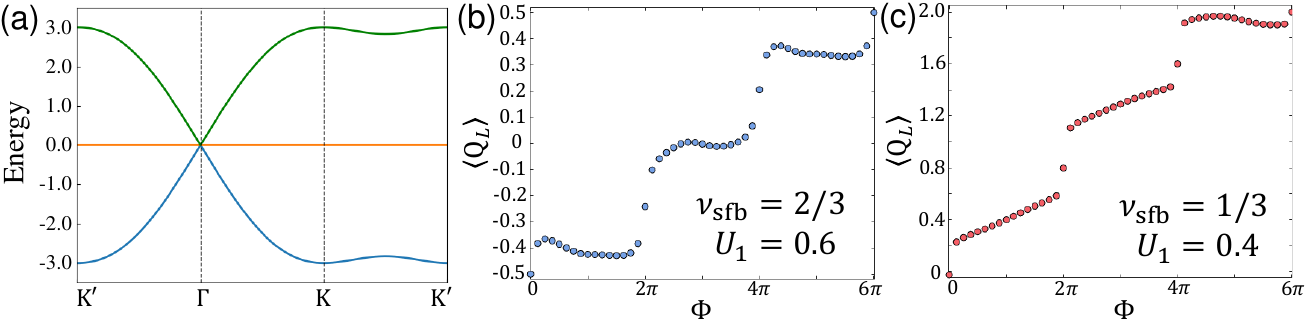}
	\caption{(a) Energy bands with linear band touching when $E_A=0$. (b, c) FQAH effects demonstrated by DMRG calculation of charge pumping at (b) $\nu_{\rm sfb}=2/3$ filling of the singular flat band and interaction strength $U_1=0.6$, or (c) $\nu_{\rm sfb}=1/3$ and interaction strength $U_1=0.4$.}
	\label{Fig:LinearTouching}
\end{figure}

Here we examine the scenario of a linear band touching by setting $E_A=0$. Importantly, we found that the FQAH phases are preserved. 
At $\nu_{\rm sfb}=2/3$ ($1/3$) filling of the singular flat band, DMRG calculations reveal that the FQAH phase can survive up to interaction strength $U_1=0.8$ ($0.7$) with fractionally quantized Hall conductivity of $e^2/3h$ ($2e^2/3h$), as shown by the results of charge pumping in Figs.~\ref{Fig:LinearTouching}(b, c). 
Compared to the results of quadratic band touching [Fig.~2(a) of the main text], the charge pumping curves in Figs.~\ref{Fig:LinearTouching}(b, c) are more oscillatory, indicating a smaller topological gap in the scenario of linear band touching.

\end{document}